\documentclass[a4paper,11pt,nontitlepage]{article}
\pdfoutput=1 
\usepackage{jheppub} 
\usepackage[T1]{fontenc} 
\RequirePackage[numbers,sort&compress]{natbib}

\usepackage[utf8]{inputenc}               
\usepackage{amsmath}
\usepackage{mathtools}
\usepackage[T1]{fontenc}                  
\usepackage{amssymb}
\usepackage{graphicx}
\usepackage{subfigure}
\usepackage{braket}
\usepackage{dsfont}
\usepackage{siunitx}
\usepackage{mathtools, slashed}
\usepackage[dvipsnames]{xcolor}
\usepackage[export]{adjustbox}
\usepackage{amsfonts}
\usepackage{amsmath}
\usepackage{amssymb}
\usepackage{array}
\usepackage{bbold}
\usepackage{bm}
\usepackage{hyperref}
\hypersetup{
    colorlinks=true,
    urlcolor=blue,
    linkcolor = blue,
    urlcolor  = blue,
    citecolor = blue,
    anchorcolor = blue,
    pdftitle={Roper-dimer},
    pdfauthor={Daniel und co.},
    pdfsubject={Roper resonance}}

\graphicspath{{pics/}}
\numberwithin{equation}{section}
\title{Particle-dimer approach for the Roper resonance in a finite volume}

\author[a]{Daniel Severt}
\author[a,b]{, Maxim Mai}
\author[a,c,d]{and Ulf-G. Mei{\ss}ner}

\affiliation[a]{Helmholtz-Institut f\"ur Strahlen- und Kernphysik (Theorie) and Bethe Center for Theoretical
  Physics, Universit\"at Bonn, D--53115 Bonn, Germany}
\affiliation[b]{Institute for Nuclear Studies and Department of Physics, The George Washington University,
  Washington, DC 20052, USA}
\affiliation[c]{Institute for Advanced Simulation, Institut f\"ur Kernphysik and J\"ulich Center for Hadron
  Physics, Forschungszentrum J\"ulich, D-52425 J\"ulich, Germany}
\affiliation[d]{Tbilisi State University, 0186 Tbilisi, Georgia}

\emailAdd{severt@hiskp.uni-bonn.de}
\emailAdd{mai@hiskp.uni-bonn.de}
\emailAdd{meissner@hiskp.uni-bonn.de}
\abstract{We propose a new finite-volume approach which implements two- and three-body dynamics in a
transparent way based on an Effective Field Theory Lagrangian. The formalism utilizes a particle-dimer
picture and formulates the quantization conditions based on the self-energy of the decaying
particle. The formalism is studied for the case of the Roper resonance, using input from lattice QCD and phenomenology. Finally, finite-volume energy eigenvalues are predicted and compared to existing results of lattice QCD calculations. This crucially provides initial guidance on the necessary level of precision for the finite-volume spectrum.
}

\begin{document}

\maketitle	
\section{Introduction}
\label{sec:intro}

Our understanding of the strong interaction is tested by our ability to unravel the pattern and
production mechanism behind its bound states and resonances. The exploration of this non-trivial and very rich spectrum
is the main motivation behind the large international experimental
programs at, e.g., MAMI (Germany), ELSA (Germany), Jefferson Laboratory (USA), Spring-8 (Japan) or
CERN (Switzerland), see
\cite{Crede:2013kia,Thiel:2022xtb,Liu:2019zoy,Guo:2017jvc,Brambilla:2019esw,Ali:2019roi,Lebed:2016hpi} for some  recent reviews.
Unravel-ing the pattern of the resonance spectrum and the mechanism behind its generation has also
prompted the develop-ment of many theoretical tools such as quark
models~\cite{Loring:2001kx,Capstick:1986bm,Capstick:1993kb}, or Dyson-Schwinger
approaches \cite{Qin:2011dd,Roberts:1994dr,Eichmann:2016yit}. While some features of the resonance
spectrum seem to be captured by such approaches, they also include some uncontrolled approximations and
do not allow for a first-principle connection to Quantum Chromodynamics (QCD). Lattice QCD provides such
an approach, which already reshaped the field of hadron spectroscopy leading to many valuable
insights on, e.g., the ground state spectrum of baryons~\cite{Durr:2008zz} and many excited
states, see e.g. Refs.~\cite{Engel:2013ig, Alexandrou:2014mka, WalkerLoud:2008bp, Bulava:2010yg, Dudek:2010wm, Alexandrou:2012rm, Alexandrou:2013fsu, Mai:2019pqr, Fischer:2020yvw, Wilson:2015dqa} as recently reviewed in
Ref.~\cite{Mai:2022eur}.

\clearpage

Two paramount examples of the puzzles in the baryon spectrum are the negative strangeness
$\Lambda(1405)1/2^-$-resonance with its double pole structure (see for example the recent
reviews~\cite{Mai:2020ltx,Hyodo:2020czb,Meissner:2020khl}) and the first excited state of the
nucleon, the Roper resonance $N(1440)1/2^+$. The latter is considerably lighter than the parity partner
of the nucleon, the $N(1535)1/2^-$. This is at odds with the  quark model
expectation~\cite{Isgur:1977ef,Isgur:1978wd}, associated there with the second radial excitation of
the nucleon. More recent phenomenological analyses revealed the complex analytic structure of the
Roper~\cite{Krehl:1999km,Arndt:2006bf, Doring:2009yv, AlvarezRuso:2010xr} including the  strong coupling
to the three-body ($\pi\pi N$) channels distorting its shape from the usual Breit-Wigner form. Ab-initio
access to such three-body systems from lattice QCD has been obscured for a long time due to computational
complexity and, equally importantly, by the lack of 
theoretical tools relating lattice results to real-world quantities. The need for such tools is simply necessitated by the fact that in lattice methodology QCD Green's functions are determined numerically in a finite volume. Ultimately, this leads to a discretization of the obtained real-valued spectrum to be related to the infinite-volume (real-world) 
interaction spectrum, where, in general, complex-valued amplitudes occur. This cannot be overcome in an adiabatic enlargement of the considered volume and mathematical mapping is required, usually referred to as the \emph{quantization condition}, for dedicated reviews see Refs.~\cite{Hansen:2019nir,Rusetsky:2019gyk,Mai:2021lwb}. 
Lattice results for such systems are becoming available, see Refs.~\cite{Lang:2016hnn,Kiratidis:2016hda,Liu:2016uzk,Horz:2019rrn,Blanton:2019vdk,Culver:2019vvu,Fischer:2020jzp,Hansen:2020otl,Alexandru:2020xqf,Blanton:2021llb,NPLQCD:2020ozd,Buhlmann:2021nsb,Mai:2021nul,Garofalo:2022pux}.
This is also partially fostered by the recent progress deriving three-body quantization conditions~\cite{Polejaeva:2012ut,Briceno:2012rv,Meissner:2014dea,Hansen:2014eka,Hansen:2015zga,Guo:2016fgl,Hammer:2017uqm,Hammer:2017kms,Briceno:2017tce,Guo:2017ism,Meng:2017jgx,Guo:2018xbv,Mai:2017bge,Mai:2018djl,Doring:2018xxx,Mai:2019fba,Guo:2019hih,Blanton:2019igq,Hansen:2020zhy,Guo:2020kph,Blanton:2020gmf,Pang:2022nim,Brett:2021wyd,Muller:2021uur,Muller:2020wjo,Hansen:2021ofl,Blanton:2021mih,Jackura:2022gib,Muller:2022oyw}.

In the present work we propose and test a new approach to the three-body quantization conditions which can
serve as a transparent approach to access resonant systems in a finite-volume. Our formalism builds on
the previous work~\cite{Severt:2020jzc} and is based on the particle-dimer
framework~\cite{Kaplan:1996nv,Bedaque:1998kg,Bedaque:1998km,Braaten:2004rn,Hammer:2017kms}, which
conveniently allows us to express the self-energy diagram of a resonant field in terms of either
ordinary (asymptotically stable) meson and baryon fields or,  alternatively, one of these fields can also be replaced by an unstable field from the particle-dimer Lagrangian. The latter in turn  acquires a complex-valued self-energy due to the coupling to stable fields going on-shell. Obviously, the interplay of these two effects leads to an on-shell configuration of three stable intermediate particles. Indeed, these are precisely the configurations which lead to power-law finite-volume effects. 
These finite-volume effects fall into two different categories: First, effects that scale as a power of $1/L$, where $L$ is the length of the cubic volume with periodic boundary conditions in which finite-volume calculations are performed, and, second, the so-called exponentially suppressed finite-volume effects. 
The latter are for example proportional to $\exp(- M_{\pi} L)$, where $M_{\pi}$ denotes the mass of the pion, i.e. the lightest asymptotic particle in QCD. 
Thus, neglecting these exponentially suppressed contributions, one can separate off
the volume-dependent from the volume-independent quantities which ultimately allows one to map finite- to
infinite-volume quantities. To demonstrate the advantages and limitations of the present work, we
concentrate specifically on the complicated Roper resonance including the $\pi N$ and $\pi\pi N$
dynamics using $\Delta$ and $\sigma$ auxiliary dimer-fields. Given the presently still scarce lattice results
in this sector~\cite{Lang:2016hnn,Kiratidis:2016hda,Liu:2016uzk} we estimate the volume-independent
quantities from phenomenology. The predicted finite-volume spectrum is then compared to lattice
results both in the two- and three-body sector~\cite{Guo:2018zss,Lang:2016hnn}.

The manuscript is organized in the following way: First, we introduce the theoretical framework in
section~\ref{sec:covariant-framework}. Then, we determine the self-energy of the Roper resonance within
our theory in section~\ref{sec:se-roper}. The sections~\ref{sec:Dimerfield} and~\ref{sec:se-dimer}
discuss the particle-dimer fields and their contributions to the Roper self-energy, respectively.
After that, the finite-volume formalism is introduced in section~\ref{sec:FVformalism}. Our numerical
calculations are discussed in section~\ref{sec:Numerics} and the results are given in~\ref{sec:results}.
Finally, we conclude with a brief summary and outlook in section~\ref{sec:summary}.

\section{Covariant non-relativistic framework}
\label{sec:covariant-framework}

We begin with an introduction of the covariant non-relativistic effective field theory, following the general
formalism of Refs.~\cite{Braaten:2004rn,Hammer:2017uqm,Hammer:2017kms,Colangelo:2006va,Bernard:2008ax},
see also Ref.~\cite{Meissner:2022odx} for a pedagogical introduction.
To describe a few-particle system containing pions ($\pi$) and nucleons ($N$), such as the three-particle
$N \pi \pi$-system, we introduce the following Lagrangian 
\begin{align}
\begin{split}
\mathcal{L}_{\pi \pi N} = \mathcal{L}_{\rm dyn} &+ c_1 \phi^{\dagger}  \phi^{\dagger} \phi \phi + c_2 \psi^{\dagger} \phi^{\dagger} \phi \psi + c_3 \psi^{\dagger} \phi^{\dagger} ( \phi^{} + \phi^{\dagger} ) \phi \psi + c_4 \psi^{\dagger} \phi^{\dagger} \phi^{\dagger} \phi^{} \phi \psi + \ldots  \; .
\end{split} \label{piN-Lagrangian}
\end{align}
Here, $\phi$ is the non-relativistic pion field and $\psi$ the non-relativistic nucleon field. The
interaction between these particles is parameterized by the low-energy constants (LECs) $c_{1,2,3,4}$.
The ellipses denote terms with higher numbers of (pion) field insertions not required for
the purpose of this work and terms with derivatives, which are not taken into account for now.
Containing short-range physics, the LECs are in general not known, but can be determined from experimental
data or lattice QCD results. The LEC $c_{1}$, for example, can be related to the $\pi \pi$ scattering
length. The dynamical part of the covariant Lagrangian for the pions and nucleons is given by~\cite{Bernard:2008ax}
\begin{align}
\begin{split}
  \mathcal{L}_{\rm dyn} &= \mathcal{L}_{\phi} + \mathcal{L}_{\psi} = \phi^{\dagger} 2 W_{\pi} \left( i \partial_t -
  W_{\pi} \right) \phi + \psi^{\dagger} 2 W_{N} \left( i \partial_t - W_{N} \right) \psi \; , 
\end{split} \label{piN-dyn-Lagrangian}
\end{align}
where
\begin{align}
W_{\pi} = \big[ M_{\pi}^2 - \vec{\nabla}_{\phantom{\pi}}^2  \big]^{1/2} \; , \quad W_{N} = \big[ m_{N}^2 - \vec{\nabla}_{\phantom{N}}^2  \big]^{1/2} \;  . 
\end{align}
The differential operators $W_{\pi}$ and $W_{N}$ contain the pion mass $M_{\pi}$ and the nucleon mass $m_{N}$,
respectively. The square root structure of these operators leads to the relativistic energy-momentum
relation in momentum space and ensures that the resulting amplitudes (e.g. two-particle scattering amplitudes)
are relativistically invariant. This is not the case in the commonly used alternative non-relativistic treatment, which uses the Schr{\"o}dinger equation to describe the dynamics of the free particles, see e.g. Ref.~\cite{Braaten:2004rn}.

The Lagrangian~\eqref{piN-Lagrangian} defines the pattern of the interactions driving the construction of various
$n$-particle scattering amplitudes. However, already the case of three particles would result in a
tremendous amount of Feynman diagrams. This is where the particle-dimer formalism becomes particularly
handy, which we, therefore, utilize to address the Roper resonance. In the particle-dimer formalism one
introduces an auxiliary field, called \emph{dimer field} (sometimes also referred to \emph{isobar},
see e.g. Ref.~\cite{Mai:2017vot}), that incorporates two-particle dynamics and scattering. This means one effectively reduces a three-body problem to a two-body problem, which can be solved with much more ease.
A common example to show the strength of the dimer formalism is the calculation of the scattering
amplitude of three identical bosons, see e.g. Ref.~\cite{Braaten:2004rn}. In this case one introduces a
dimer field, which describes the two-particle scattering of these bosons. Then, to obtain the three-particle
scattering amplitude, one calculates the scattering of one boson with the dimer field, which is equivalent
to three-particle scattering. The validity of this formalism has been discussed already several times in
the literature, see e.g. Refs.~\cite{Braaten:2004rn,Hammer:2017kms,Hammer:2017uqm,Muller:2020vtt,Muller:2020wjo}.
However, the situation becomes more complex if one has three non-identical particles, like in our case with
nucleons and pions. To investigate the Roper resonance in the $N \pi \pi$-system, we need to introduce
three different dimer fields. The first dimer field is the $\Delta (1232)$ resonance (from here on called
the $\Delta$) with quantum numbers $J^P = 3/2^{+}$. This dimer field takes into account intermediate P-wave
nucleon-pion interactions  and its quantum numbers together with a pion overlap with the Roper resonance.
The second dimer field is the $\sigma$ with the quantum numbers $J^P = 0^{+}$, i.e. the scalar-isoscalar
resonance $f_0 (500)$, formerly known as the $\sigma$-meson. It accounts for intermediate S-wave pion-pion
interactions. Also here the quantum numbers of the $f_0 (500)$ with a nucleon can have an overlap with the
Roper. Finally, the third dimer field $R$ is for the Roper resonance itself, which has the quantum numbers
of the nucleon ($J^P = 1/2^{+}$) but a larger mass\footnote{Note that there is in principle also a nucleon pole appearing in the Roper system, due to the identical quantum numbers. This pole has to be taken into account in a lattice QCD calculations, for a recent example of a lattice calculation of a 3-point function see Ref.~\cite{Barca:2022uhi}. However, in our work we look at energies larger than the nucleon mass so that an explicit inclusion is not necessary.}. Considering all above dimer-fields,
the particle-dimer Lagrangian takes the form 
\begin{align}
\begin{split}
\mathcal{L}_{\rm Dimer} = \mathcal{L}_{\rm dyn} + \mathcal{L}_{T} \; , \label{Dimer-Lagrangian}
\end{split}
\end{align}
where the dimer fields and their interactions are contained in $\mathcal{L}_{T}$, which reads 
\begin{align} 
\label{Dimer-T-Lagrangian}
\begin{split}
\mathcal{L}_{T} &= R^{\dagger} 2 W_{R} \left( i \partial_t - W_{R} \right) R
+ \alpha_{\Delta}^{} m_{\Delta}^2 \Delta^{\dagger} \Delta
+ \alpha_{\sigma}^{} M_{\sigma}^2 \sigma^{\dagger} \sigma\\ 
& \phantom{=}  + f_1 R^{\dagger} \phi^{\dagger} \phi^{} R - f_2 [ R^{\dagger} \phi^{} \psi^{} + R^{} \phi^{\dagger} \psi^{\dagger} ]   - f_3 [ R^{\dagger} \phi \Delta + \Delta^{\dagger} \phi^{\dagger} R ] - f_4 [ R^{\dagger} \sigma \psi + \psi^{\dagger} \sigma^{\dagger} R ] \\ 
& \phantom{=}  + g_1 \Delta^{\dagger} \phi^{\dagger} \phi^{} \Delta - g_2 [ \Delta^{\dagger} \phi^{} \psi^{} + \Delta^{} \phi^{\dagger} \psi^{\dagger} ] 
+h_1 \psi^{\dagger} \sigma^{\dagger} \sigma \psi - h_2 [ \sigma^{\dagger} \phi \phi + \sigma^{} \phi^{\dagger} \phi^{\dagger} ] \\ 
& \phantom{=} - G_{R \sigma} [ R^{\dagger} \phi^{\dagger} \sigma \psi + \psi^{\dagger} \sigma^{\dagger} \phi R ] - G_{R \Delta} [ R^{\dagger} \phi^{\dagger} \phi \Delta + \Delta^{\dagger} \phi^{\dagger} \phi R ]  - G_{\Delta \sigma} [ \Delta^{\dagger} \phi^{\dagger} \sigma \psi + \psi^{\dagger} \sigma^{\dagger} \phi \Delta ] \; .
\end{split}
\end{align} 
An important detail to note is that the dimer fields $\Delta$ and $\sigma$ are not dynamical, i.e. the Lagrangian
does not contain time or spatial derivatives of these fields. For the Roper resonance, on the other hand, the
same dynamical Lagrangian as for the nucleon and pion is introduced with $W_{R} = [ m_{R0}^2 -
  \vec{\nabla}_{\phantom{R}}^2  ]^{1/2}$ for the bare mass of the Roper $m_{R0}$. Making the Roper resonance
dynamical should give a more accurate treatment of its properties. Overall, the dimer fields are auxiliary
fields and the choice of their kinetic energy term should depend on the overall goal of the calculation.
Naturally, the introduction of derivative terms for the dimer fields results in more complex calculations,
since these terms will enter the dimer propagators. Therefore, to simplify our analysis, we keep the
$\Delta$- and $\sigma$-dimer static. Additionally, it should be stressed that the Lagrangian in
Eq.~\eqref{Dimer-T-Lagrangian} does not posess any spin- or isospin-structure. Also here the Lagrangian
can be modified to include these effects, but we do not consider them for now in this pioneering work.

There are several coupling constants in Eq.~\eqref{Dimer-T-Lagrangian} accompanying the terms describing
the interactions between the particles and dimer fields. The LECs $f_{1,2,3,4}$, $g_{1,2}$, $h_{1,2}$ and
$G_{R \sigma, R \Delta, \Delta \sigma}$ can be related to the LECs in Eq.~\eqref{piN-Lagrangian} after
integrating out the dimer fields. We also have two real mass scales $m_{\Delta}$ and $M_{\sigma}$
for the $\Delta$- and $\sigma$-dimer, respectively. Very often in the literature these mass scales
are absorbed inside the definition of the auxiliary dimer fields. We, on the other hand, want to make
sure that all appearing fields have the same dimension and later use the physical masses for
our numerical calculations.  Both of these mass scales come with prefactors 
\begin{align}
\alpha_{\Delta} = \pm 1 \; , \quad \alpha_{\sigma}= \pm 1 \; ,
\label{eq:Dimer-L}
\end{align}
which depend on the signs of the corresponding LECs in Eq.~\eqref{piN-Lagrangian}. For example,
integrating out the $\sigma$-field yields 
\begin{align}
c_{1} = - \frac{h_{2}^2}{\alpha_{\sigma}^{} M_{\sigma}^2} \; .
\end{align}
It can be seen that the sign of $c_1$ dictates the value of $\alpha_{\sigma}$, since $h_{2}^2/M_{\sigma}^2$ is
a positive number. Later in the manuscript, we will see how $\pi \pi$ scattering information
(e.g. the S-wave scattering length or the corresponding phase shifts) determine this LEC.

Another notable difference between Eq.~\eqref{Dimer-T-Lagrangian} and most other Lagrangians in the
particle-dimer picture are the interactions among the dimer fields. The Roper dimer $R$ is allowed to
decay in one of the other dimer fields, i.e. $R$ can decay into $\sigma N$, or $\Delta\pi$ pairs through the interactions proportional to $f_3$ and $f_4$, respectively. An example for a particle-dimer theory with two dimer
fields that can interact with each other can be found in Ref.~\cite{Pang:2020pkl}. After integrating
out the dimer fields, interactions with an odd number of pion fields can be obtained, e.g. the term
proportional to $c_3$ in Eq.~\eqref{piN-Lagrangian} can change the number of particles. This yields
the feature that a two-particle $N \pi$ initial state could result in a three-particle $N \pi \pi$ final
state and vice versa. Obviously, this then also means that there can in principle be a four-particle
$N \pi \pi \pi$ final state when starting with an initial three-particle $N \pi \pi$ state, etc..
However, in practice we avoid such a four-particle (and higher) final state by a suitable energy/momentum cutoff.

The particle-dimer Lagrangian Eq.~\eqref{Dimer-T-Lagrangian} yields the following Feynman rules for the propagators: 
\begin{align}
-i S_{N} \left( p_0, \vec{p} \right) = \frac{i}{2 \omega_{N} (\vec{p}) \left[ p_0 -
\omega_{N} (\vec{p}) + i \epsilon \right]} \; , \quad \omega_{N} (\vec{p}) = \sqrt{| \vec{p} |^{2} + m_{N}^{2}} \; ,
\label{nucl-prop}
\end{align}
and
\begin{align}
-i S_{\pi} \left( p_0, \vec{p} \right) = \frac{i}{2 \omega_{\pi} (\vec{p}) \left[ p_0 - \omega_{\pi} (\vec{p})
+ i \epsilon \right]} \; , \quad \omega_{\pi} (\vec{p}) = \sqrt{| \vec{p} |^{2} + M_{\pi}^{2}} \; . \label{pion-prop}
\end{align}
Looking at $\omega_{N}$ and $\omega_{\pi}$, one notes that the square root differential operator in the
dynamical part of the Lagrangian leads to the well-known energy-momentum relation. Our notation for the
propagators follows the common sign convention used in the literature, see e.g.~\cite{Bernard:2008ax}.
For the dimer fields, we have the bare propagator of the Roper resonance  
\begin{align}
-i S_{R} \left( p_0, \vec{p} \right) = \frac{i}{2 \omega_{R} (\vec{p}) \left[ p_0 - \omega_{R} (\vec{p})
+ i \epsilon \right]} \; , \quad \omega_{R} (\vec{p}) = \sqrt{| \vec{p} |^{2} + m_{R0}^{2}} \; , \label{roper-prop}
\end{align}
and the bare $\Delta$ and $\sigma$ propagators 
\begin{align}
-i D_{\Delta}^0 \left( p_0, \vec{p} \right) = \frac{i}{\alpha_{\Delta}^{} m_{\Delta}^2} \; , \quad -i D_{\sigma}^0
\left( p_0, \vec{p} \right) = \frac{i}{\alpha_{\sigma}^{} M_{\sigma}^2} \; , \label{bare-dimer-props}
\end{align}
where the latter are constant with respect to the particle energy. An explicit momentum dependence can be
given to $D_{\Delta}^0$ and $D_{\sigma}^0$ by either adding higher order terms in the particle-dimer
Lagrangian Eq.~\eqref{Dimer-T-Lagrangian} or by ``dressing'' the propagators with the respective dimer self-energies.
The latter is discussed in detail in section~\ref{sec:Dimerfield}.

\section{Self-energy of the Roper resonance}
\label{sec:se-roper}

The dressed propagator of the Roper resonance is given by 
\begin{align}
S_{R}^{d} \left( p_0, \vec{p} \right) = \frac{1}{2 \omega_{R} (\vec{p}) \left[ \omega_{R} (\vec{p}) - p_0 - i \epsilon \right] - \Sigma_{R} (p_0, \vec{p}) } \; , \label{dressed-roper-prop}
\end{align}
where $\Sigma_{R} (p_0, \vec{p})$ is Roper self-energy. The pole of the propagator is obtained by finding
the zeros of the denominator, i.e. 
\begin{align}
2 \omega_{R} (\vec{p}) \left[ \omega_{R} (\vec{p}) - p_0 \right] - \Sigma_{R} (p_0, \vec{p}) = 0 \; . 
\end{align}
In the infinite volume, one possibility to parameterize the pole is to choose the rest-frame, $\vec{p}=0$,
and set $p_0 = z$ for $z = m_R - i\Gamma_R/2$,  with $m_R$ the physical mass of the Roper resonance and
$\Gamma_R$ its width. The equation for the pole then reads 
\begin{align}
2 m_{R0} \left[ m_{R0} - z \right] - \Sigma_{R} (z, \vec{0}) = 0 \; ,  
\end{align}
which can be reordered to give
\begin{align}
z = m_{R0} - \frac{1}{2 m_{R0} } \Sigma_{R} (z) = m_{R0} - \frac{1}{2 m_{R0} } \Big( \text{Re}
\left\lbrace \Sigma_{R} (z) \right\rbrace + i \text{Im} \left\lbrace \Sigma_{R} (z) \right\rbrace \Big)  \; ,  
\end{align}
where the self-energy has been separated into its real and imaginary part. It is then straightforward to
identify the physical mass and width 
\begin{align}
m_R = m_{R0} - \frac{1}{2 m_{R0} } \text{Re} \left\lbrace \Sigma_{R} (z) \right\rbrace \; , \quad \text{and}
\quad \Gamma_R = \frac{1}{m_{R0} } \text{Im} \left\lbrace \Sigma_{R} (z) \right\rbrace \; . \label{def-mass-and-width}
\end{align}
These two relations can, of course, only be solved iteratively, since the self-energy depends on $z$ itself.
If the imaginary part of the self-energy vanishes, the width $\Gamma_R$ is zero. A vanishing real part,
on the other hand, allows to set the bare mass equal to the physical mass, i.e. $m_R = m_{R0}$.

\begin{figure}[t]
	\begin{center}  
		\includegraphics[width=1.0\linewidth]{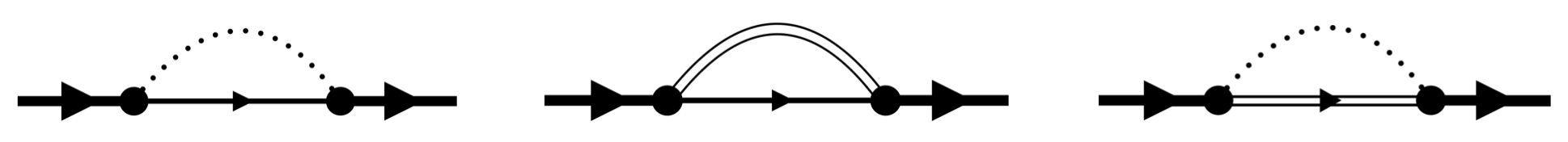}
	\end{center}
	\caption{
	  Feynman diagrams contributing to the Roper resonance mass at one-loop order. The thick solid
          line with an arrow, the solid line with an arrow and the double solid line with an arrow refer
          to the Roper resonance, the nucleon, and the $\Delta$-dimer field, respectively. The dotted line
          represents pions and the double solid line the $\sigma$-dimer fields.    
	} 
	\label{fig:RoperSE-loops}
\end{figure}

Looking at the full particle-dimer Lagrangian in Eq.~\eqref{Dimer-T-Lagrangian}, we see that there are
several interactions which lead to different contributions to the self-energy, as depicted in
Fig.~\ref{fig:RoperSE-loops}. At one-loop order, the first option is a pion and a nucleon inside the loop.
Since both are stable particles and we know that the Roper $R$ can decay into a $N \pi$ final-state, we
expect this diagram to be of great importance. The next option is $N$ and the $\sigma$-dimer inside the loop.
This diagram is interesting, because the dimer itself is an unstable particle. We know that the Roper
can decay into the $N \sigma$ pair, but we expect that the $\sigma$ decays further into two pions,
which would leave us with the three particle ($N \pi \pi$) final-state. This is similar to the third option,
a $\pi$ and $\Delta$-dimer inside the loop. Also here, the $\Delta$ can decay further into a $N \pi$ state,
which again results in a three-particle $N \pi \pi $-system. Note, further that  one-loop tadpole diagrams
do not appear in the non-relativistic theory.
We can summarize these statements, see Fig.~\ref{fig:RoperSE-loops}, into the following equation
\begin{align}
\Sigma_{R} (p_0, \vec{p}) = \Sigma_{N \pi} (p_0, \vec{p}) + \Sigma_{N \sigma} (p_0, \vec{p})
+ \Sigma_{\Delta \pi} (p_0, \vec{p}) \; , \label{Roper-Self-Energy}
\end{align}
and our goal is to calculate the different self-energy contributions.

We start by evaluating the self-energy $\Sigma_{N \pi} (p_0, \vec{p}) $. Applying the Feynman rules, we obtain 
\begin{align}
\begin{split}
i \Sigma_{N \pi} \left( p \right) &= \int \frac{d^4 k}{(2 \pi)^4} (- i f_2)^2 \left[ -i S_{N} (p-k) \right]
\left[ -i S_{\pi} (k) \right] 
= f_2^2 \int \frac{d^4 k}{(2 \pi)^4} S_{N} (p-k) S_{\pi} (k) \; . 
\end{split}
\end{align} 
From here on, we use the four-vector $p$ as a shorthand notation for $(p_0, \vec{p})$. After dividing
by $i$ on both sides we find
\begin{align}
\Sigma_{N \pi} \left( p \right) = f_2^2 J_{N \pi} \left( p \right) \; , \label{self-energy-pi-N-1}
\end{align}
with
\begin{align}
\begin{split}
J_{N \pi} \left( p \right) = \int \frac{d^4 k}{(2 \pi)^4 i} \, \frac{1}{2 \omega_{N} (\vec{p} - \vec{k}) \big[ \omega_{N}(\vec{p} - \vec{k}) - (p_0 - k_0) - i \epsilon \big]}  \frac{1}{2 \omega_{\pi} (\vec{k}) \big[ \omega_{\pi} (\vec{k}) - k_0 - i \epsilon \big]} \, . \label{J-pi-N}
\end{split}
\end{align}
This is the main one-loop scalar integral appearing in the covariant non-relativistic frame-work. The evaluation
of this integral is non-trivial, due to the square root structures appearing in the denominator,
see e.g. Ref.~\cite{Gasser:2011ju}. However, the first step is straightforward, integrating over
the time component of the loop momentum $k_0$, i.e.  
\begin{align}
\begin{split}
  J_{N \pi} \left( p \right) & = \int_{-\infty}^{+\infty} \frac{d k_0}{2 \pi i} \int \frac{d^3 k}{(2 \pi)^3} \,
  \frac{1}{4 \omega_{N} (\vec{p} - \vec{k}) \omega_{\pi} (\vec{k})} \\
  & \phantom{= \int_{-\infty}^{+\infty} \frac{d k_0}{2 \pi i} \int \frac{d^3 k}{(2 \pi)^3} } \; \times \left \lbrace
  \frac{1}{ \big[ \omega_{N}(\vec{p} - \vec{k}) - (p_0 - k_0) - i \epsilon \big] \big[ \omega_{\pi} (\vec{k}) - k_0
      - i \epsilon \big] } \right \rbrace \; . 
\end{split}
\end{align}
Looking at the denominator inside the brackets, we see that it has two poles in the complex $k_0$-plane, namely one
in the upper half (positive imaginary part) and one in the lower half (negative imaginary part).
Using Cauchy's theorem, we can solve the integral by calculating a contour integral around one of the poles.
Choosing a contour around the upper pole\footnote{The result of the integral does not change, if one
  would choose the pole in the lower half.}, we obtain 
\begin{align}
\begin{split}
J_{N \pi} \left( p \right) = \int \frac{d^3 k}{(2 \pi)^3} \, \frac{1}{4 \omega_{N} (\vec{p} - \vec{k}) \omega_{\pi}
 (\vec{k}) \big[ \omega_{N}(\vec{p} - \vec{k}) + \omega_{\pi} (\vec{k}) - p_0 - i \epsilon \big] } \; .
\label{J-pi-N-2} 
\end{split}
\end{align}
We are left with a three-dimensional integral over the spatial momentum components, which will also be our
starting point when we consider the finite-volume case later in section~\ref{sec:FVformalism}. One observes
that the integral has a pole for $p_0>0$ (taking $\epsilon \to 0$) and that the integral is logarithmically
divergent. It is therefore practical to use dimensional regularization for the further evaluation.
In $D$ dimensions Eq.~\eqref{J-pi-N-2} takes the form 
\begin{align}
\begin{split}
J_{N \pi} \left( p \right) = \int \frac{d^D k}{(2 \pi)^D} \, \frac{1}{4 \omega_{N} (\vec{p} - \vec{k})
\omega_{\pi} (\vec{k}) \big[ \omega_{N}(\vec{p} - \vec{k}) + \omega_{\pi} (\vec{k}) - p_0 - i \epsilon \big] } \; .
\end{split}
\end{align}
The main complexity still comes from the square root terms in the denominator. To simplify matters, let us
consider the same integral in the rest frame, i.e. $p=(E, \vec{0})$,
\begin{align}
\begin{split}
J_{N \pi} \left( E \right) = \int \frac{d^D k}{(2 \pi)^D} \, \frac{1}{4 \omega_{N} ( \vec{k}) \omega_{\pi} (\vec{k}) \big[ \omega_{N}( \vec{k}) + \omega_{\pi} (\vec{k}) - E - i \epsilon \big] } \; ,
\end{split}
\end{align}
such that we can rewrite the integrand as
\begin{align}
\begin{split}
	\frac{1}{4 \omega_{N} ( \vec{k}) \omega_{\pi} (\vec{k}) \big[ \omega_{N}( \vec{k}) + \omega_{\pi} (\vec{k}) - E - i \epsilon \big] } &= \frac{1}{2 E} \, \frac{1}{ |\vec{k}|^2 - q^2 (E) - i \epsilon' } \\
	& \phantom{=} + \frac{1}{4 \omega_{N} ( \vec{k}) \omega_{\pi} (\vec{k}) \big[ \omega_{N}( \vec{k}) + \omega_{\pi} (\vec{k}) + E + i \epsilon \big] } \\
	& \phantom{=} + \frac{1}{4 \omega_{N} ( \vec{k}) \omega_{\pi} (\vec{k}) \big[ \omega_{N}( \vec{k}) - \omega_{\pi} (\vec{k}) - E + i \epsilon \big] } \\
	& \phantom{=} + \frac{1}{4 \omega_{N} ( \vec{k}) \omega_{\pi} (\vec{k}) \big[ - \omega_{N}( \vec{k}) + \omega_{\pi} (\vec{k}) - E + i \epsilon \big] } \; , \label{integrand-rewritten}
\end{split}
\end{align}
with 
\begin{align}
	q^2 (E) = \frac{\lambda \left( E^2, m_{N}^2 , M_{\pi}^2 \right)}{4 E^2} \; , 
\end{align}
where we used the  K{\"a}ll{\'e}n triangle function $\lambda(x,y,z) = x^2 + y^2 +z^2 - 2xy - 2xz -2yz$. The
rearrangement of the integrand allows us to isolate the pole of the quotient, $q^2(E)$, which can be seen
in the first term on the  right-hand side of Eq.~\eqref{integrand-rewritten}. The remaining three terms on
the right-hand side are regular, which means that they do not contain a pole anymore for physical values of $E$. Note that in this work we consider energies above the nucleon mass. Therefore,
these terms can be expanded in powers of the integration momentum $\vec{k}$ leading to polynomials
in $|\vec{k}|$ which vanish in dimensional regularization. We are left with 
\begin{align}
J_{N \pi} \left( E \right) = \frac{1}{2 E} \int \frac{d^D k}{(2 \pi)^D} \, \frac{1}{ |\vec{k}|^2 - q^2 (E)
  - i \epsilon' } \; , 
\end{align}
which is evaluated with standard methods. After taking the limit $D \rightarrow 3$ we obtain 
\begin{align}
  J_{N \pi} \left( E \right) = - \frac{1}{8 \pi E} \sqrt{-q^2(E) - i \epsilon'} = \frac{i \lambda^{1/2}
    \left( E^2, m_{N}^2 , M_{\pi}^2 \right) }{16 \pi E^2}  \; , \label{J-pi-N-result-restframe} 
\end{align}
where we used that $\lim_{\epsilon' \rightarrow 0} \sqrt{-q^2(E) \pm i \epsilon'} = \pm i q(E)$. The result of
Eq.~\eqref{J-pi-N-result-restframe} in an arbitrary reference frame reads~\cite{Gasser:2011ju}
\begin{align}
  J_{N \pi} \left( p \right) = \frac{i \lambda^{1/2} \left( p^2, m_{N}^2 , M_{\pi}^2 \right) }{16 \pi p^2}
  = \frac{i \lambda^{1/2} \left( s, m_{N}^2 , M_{\pi}^2 \right) }{16 \pi s}  \; , \label{J-pi-N-result-full} 
\end{align}
with $s=p^2 = p_0^2 - | \vec{p} |^2$ the usual Mandelstam variable. Thus, the self-energy of the Roper
resonance becomes 
\begin{align}
\Sigma_{N \pi} \left( p \right) = \frac{i f_2^2 }{16 \pi p^2} \lambda^{1/2} \left( p^2, m_{N}^2 , M_{\pi}^2 \right) \; ,
\label{self-energy-pi-N-2}
\end{align}
which is a notable result. Specifically, the function $J_{N \pi}$ and with it  the Roper self-energy is
purely imaginary at the energies of interest, i.e. $p^2\approx m_R^2$.

Next, we consider the self-energy contributions with dimer fields, i.e. $\sigma N$ and $\Delta\pi$
loop-diagram contributions. Taking $\Sigma_{\Delta \pi}$ as an example, we obtain
\begin{align}
\begin{split}
i \Sigma_{\Delta \pi} (p) = f_3^2 \int \frac{d^4 k}{(2 \pi)^4} D_{\Delta}^0 (p-k) S_{\pi} (k) 
= - \frac{f_3^2}{\alpha_{\Delta}^{} m_{\Delta}^2}  \int \frac{d^4 k}{(2 \pi)^4} \frac{1}{2 \omega_{\pi} (\vec{k})
\big[ \omega_{\pi} (\vec{k}) - k_0 - i \epsilon \big]} \; , 
\end{split}
\end{align}
which is basically a tadpole integral, i.e. an integral over a single propagator, due to the constant
$D_{\Delta}^0$ propagator. These tadpole diagrams usually do not exist in non-relativistic EFTs, since
they vanish within time-ordered perturbation theory. However, if such a diagram shows up, a common way
to treat the $k_0$-integral is to rewrite it as a contour integral according to Cauchy's theorem. 
For example, one can evaluate the $k_0$-integral by choosing the contour in the upper $k_0$-plane excluding
the pole. Then, $\Sigma_{\Delta \pi} (p)$ vanishes like the other tadpole contributions, which is the
usual procedure, see Ref.~\cite{Meissner:2022odx} and the references therein for more information. On the
other hand, if one would decide to include the pole (lower plane), the following would happen: The $k_0$-integral
is replaced by $2 \pi i $ and a spatial $\vec{k}$-integral over $1/\omega_{\pi} (\vec{k})$ remains. But
this expression does not posses a pole and, thus, one can expand the denominator in powers of the momentum
$|\vec{k}|$, like before, to obtain a polynomial. Dimensional regularization is then used to make the
polynomial terms disappear, so that again $\Sigma_{\Delta \pi} (p) = 0$. This illustrates that the loop
integral vanishes no matter how the $k_0$-integral is performed. An analogous calculation for the
$N \sigma$-case shows that also $\Sigma_{N \sigma} (p)=0$.

This of course cannot be the final answer, which roots in the fact that the dimer propagators are not
dynamical, see Eq.~\eqref{Dimer-T-Lagrangian}. Interestingly, and as we will discuss below, improving this
by dressing dimer propagators actually introduces three-particle dynamics in the intermediate states.

\section{Dressed dimer fields}
\label{sec:Dimerfield}

We have seen that a constant dimer propagator leads to a vanishing particle-dimer self-energy. Obviously,
the constant propagator is just a first approximation and higher correc-tions have to be taken into account.
To do this, we consider the self-energies of the dimer fields and \emph{dress} the propagators as  
\begin{align}
D_{\Delta} \left( p \right) = - \frac{1}{\alpha_{\Delta}^{} m_{\Delta}^2 + \Sigma_{\Delta}^{} (p) } \; ,
\label{dressed-delta-prop}
\end{align}
for the $\Delta$-dimer propagator and 
\begin{align}
D_{\sigma} \left( p \right) = - \frac{1}{\alpha_{\sigma}^{} M_{\sigma}^2 + \Sigma_{\sigma}^{} (p) } \; ,
\label{dressed-sigma-prop}
\end{align}
for the $\sigma$-dimer. The self-energies $\Sigma_{\Delta}$ and $\Sigma_{\sigma}$ are given by 
\begin{align}
\Sigma_{\Delta}^{} \left( p \right)  = g_2^2 \int \frac{d^4 k}{(2 \pi)^4 i} S_{N} (p-k) S_{\pi} (k)
= \frac{i g_2^2 }{16 \pi p^2} \lambda^{1/2} \left( p^2, m_{N}^2 , M_{\pi}^2 \right) \; ,  \label{delta-self-energy}
\end{align}
and 
\begin{align}
\Sigma_{\sigma}^{} \left( p \right)  =  \frac{1}{2} (2 h_2)^2 \int \frac{d^4 k}{(2 \pi)^4 i} S_{\pi} (p-k) S_{\pi} (k)
= \frac{i h_2^2 }{8 \pi p^2} \lambda^{1/2} \left( p^2, M_{\pi}^2 , M_{\pi}^2 \right) \; ,  \label{sigma-self-energy}  
\end{align}
respectively. Note the additional symmetry factor of $1/2$ in front of the $\sigma$ self-energy. The evaluation
of these self-energies is analogous to the proof of Eq.~\eqref{self-energy-pi-N-2} in the last section. Due
to the simpler structure of the K{\"a}ll{\'e}n function in the case of two equal masses, i.e. 
\begin{align}
\lambda \left( p^2, M_{\pi}^2 , M_{\pi}^2 \right) = p^2 \left( p^2 - 4 M_{\pi}^2 \right) \; , 
\end{align}
we proceed with the $\sigma$-dimer propagator. We start by reformulating the dressed propagator as
\begin{align}
D_{\sigma} \left( p \right) = - \frac{1}{\alpha_{\sigma}^{} M_{\sigma}^2 + i c \lambda^{1/2}
\left( p^2, M_{\pi}^2 , M_{\pi}^2 \right) / p^2 } \; , \qquad c = \frac{h_2^2}{8 \pi}  \; .
\end{align}
Subsequently, we simplify the denominator by expanding the above expression so that 
\begin{align}
\label{eq:sidma-prop1}
D_{\sigma} \left( p \right) = - \frac{\alpha_{\sigma}^{} M_{\sigma}^2 p^4 - i c p^2 \lambda^{1/2} \left( p^2, M_{\pi}^2 , M_{\pi}^2 \right) }{ \alpha_{\sigma}^{2} M_{\sigma}^4 p^4 + c^2 \left( p^4 - 4 M_{\pi}^2 p^2 \right) } \; . 
\end{align}
From our initial definitions we know that $\alpha_{\sigma}^{2}=1$ and we can rewrite the denominator as
\begin{align}
\label{eq:sigma-denominator}
\alpha_{\sigma}^{2} M_{\sigma}^4 p^4 + c^2 \left( p^4 - 4 M_{\pi}^2 p^2 \right) = \left( M_{\sigma}^4 + c^2 \right)
\left( p^2 - \mu_{\sigma}^2 \right) p^2 \; ,
\end{align}
where we have introduced a new mass parameter 
\begin{align}
\mu_{\sigma}^2 = \frac{4 M_{\pi}^2 c^2}{M_{\sigma}^4 + c^2} \; . \label{mu-sigma} 
\end{align}
From Eq.~\eqref{eq:sigma-denominator}  it is evident that $\mu_{\sigma}$ is, indeed, one of the poles of
the $\sigma$-dimer. Coming back to Eq.~\eqref{eq:sidma-prop1}, we can split up the expression into a
real and an imaginary part 
\begin{align}
D_{\sigma} \left( p \right) = - \frac{\alpha_{\sigma}^{} M_{\sigma}^2}{M_{\sigma}^4 + c^2} \,  \frac{p^2}{p^2 - \mu_{\sigma}^2} + \frac{i c}{M_{\sigma}^4 + c^2} \, \frac{\lambda^{1/2} \left( p^2, M_{\pi}^2 , M_{\pi}^2 \right)}{p^2 - \mu_{\sigma}^2} \; . \label{D-sigma-split-up}
\end{align} 
In this form, we observe that $D_{\sigma} \left( p \right)$ possesses an imaginary part above the
two-particle threshold, i.e. for $p^2 = s > 4 M_{\pi}^2$. Below threshold, $D_{\sigma} \left( p \right)$ is a
real-valued function. This is in perfect agreement with the general properties of scattering amplitudes,
which in this case ($\pi\pi\to\pi\pi$ scattering) is simply proportional to the dimer propagator 
\begin{align}
T_{\pi \pi \rightarrow \pi \pi} (s) \propto D_{\sigma} \left( s \right) \; , 
\end{align}
see, e.g., Ref.~\cite{Bedaque:1998kg,Mai:2022eur}. This relation allows us to connect the coefficients
appearing in $D_{\sigma} \left( s \right)$ with observables from $\pi \pi$-scattering. The first quantity one
can look at is the scattering length $a$ defined via an effective range expansion\footnote{Note that the
sign in front of the $1/a$ term varies in the literature depending on the definition of the effective range
expansion.} 
\begin{align}
\left| \vec{q} \right| \text{cot} \delta  \left( s \right)  = + \frac{1}{a} + \mathcal{O} \big( \left| \vec{q}
\right|^2 \big) \; ,    \label{def-scattering-length} 
\end{align}
where
\begin{align}
\cot \delta  \left( s \right) = \frac{ \text{Re} \left\lbrace T_{\pi \pi \rightarrow \pi \pi} (s) \right\rbrace }{\text{Im} \left\lbrace T_{\pi \pi \rightarrow \pi \pi} (s) \right\rbrace} \; . 
\end{align}
Here, $\delta  \left( s \right)$ is the phase shift and $\vec{q}$ is the center-of-mass (CMS) three-momentum
above threshold. It can be deduced that $|\vec{q}| = \sqrt{s - 4 M_{\pi}^2} /2$. To calculate the cotangent
of the phase shift, we use the proportionality between the $\pi \pi$-scattering amplitude and the
$\sigma$-dimer propagator. We find 
\begin{align}
\frac{ \text{Re} \left\lbrace T_{\pi \pi \rightarrow \pi \pi} (s) \right\rbrace }{\text{Im} \left\lbrace T_{\pi \pi \rightarrow \pi \pi} (s) \right\rbrace} = \frac{ \text{Re} \left\lbrace D_{\sigma} \left( s \right) \right\rbrace }{\text{Im} \left\lbrace D_{\sigma} \left( s \right) \right\rbrace} = - \frac{\alpha_{\sigma}^{} M_{\sigma}^2}{c} \frac{s}{\lambda^{1/2} \left( s, M_{\pi}^2 , M_{\pi}^2 \right)} \; , 
\end{align}
and we can simplify the triangle function to $\lambda^{1/2} \left( s, M_{\pi}^2 , M_{\pi}^2 \right) =
2 \sqrt{s} |\vec{q}| $. Utilizing these identities, we obtain 
\begin{align}
\left| \vec{q} \right| \text{cot} \delta  \left( s \right) = - \frac{\alpha_{\sigma}^{} M_{\sigma}^2}{c} \frac{\sqrt{s}}{2} = - \frac{\alpha_{\sigma}^{} M_{\sigma}^2}{c} \sqrt{\left| \vec{q} \right|^2 + M_{\pi}^2 } = - \frac{\alpha_{\sigma}^{} M_{\sigma}^2 M_{\pi}^{} }{c} + \mathcal{O} \big( \left| \vec{q} \right|^2 \big)  \; . 
\end{align}
A comparison with Eq.~\eqref{def-scattering-length} shows that the scattering length $a$ is given by 
\begin{align}
a = - \frac{c}{\alpha_{\sigma}^{} M_{\sigma}^2 M_{\pi}^{}} \;  \Leftrightarrow \; a M_{\pi}^{}
= - \frac{c}{\alpha_{\sigma}^{} M_{\sigma}^2 } = - \frac{h_2^2}{8 \pi \alpha_{\sigma}^{} M_{\sigma}^2 } \; .
\label{scattering-length-dimer-constants}
\end{align}
This is a very useful result, because it fixes the ratio $h_2^2/M_{\sigma}^2$ and the value for $\alpha_{\sigma}$.
If the scattering length is positive (attractive interaction) then we must set $\alpha_{\sigma}=-1$, since all
other constants in Eq.~\eqref{scattering-length-dimer-constants} are positive. Analogously, we set
$\alpha_{\sigma}=+1$ for $a<0$ (repulsive interaction). The $\pi \pi$-scattering length in the isospin
$I=0$ channel, where the $\sigma$ resonance appears, is measured to be $a^{I=0} M_{\pi} = 0.2220
\pm 0.0128 (\text{stat.}) \pm 0.0050 (\text{syst.}) \pm 0.0037 (\text{th.})$, see Ref.~\cite{NA482:2010dug}.
Therefore, we conclude that $\alpha_{\sigma}$ must be $-1$, leading to an attractive interaction that
produces the $\sigma$ resonance.

Instead of using the scattering length to fix the LECs of the particle-dimer Lagrangian, one can also
fit them directly to the phase shifts $\delta \left( s \right)$. It is convenient to use the tangent of
$\delta (s)$ for this 
\begin{align}
\tan \delta  \left( s \right) = \frac{ \text{Im} \left\lbrace D_{\sigma} \left( s \right) \right\rbrace }{\text{Re} \left\lbrace D_{\sigma} \left( s \right) \right\rbrace} = - \frac{c}{\alpha_{\sigma}^{} M_{\sigma}^2} \frac{\lambda^{1/2} \left( s, M_{\pi}^2 , M_{\pi}^2 \right)}{s} =  a M_{\pi} \sqrt{1- \frac{4 M_{\pi}^2 }{s}}  \; .   
\end{align}
We can see that the function $\tan \delta \left( s \right)$ is zero at the threshold ($s=4 M_{\pi}^2$) and reaches
$a M_{\pi}$ for $s \to \infty$. Therefore, we expect that the above function is only able to describe the phase
shift in the low-energy region. However, this does not come as a surprise, since the $\sigma$-dimer field
is a constant at leading order, constructed specifically to approximate the low-energy regime. Another
method to calculate the parameters of the $\sigma$-dimer is to use mass and decay width of the $\sigma$-resonance.
Here, one assumes that the $\sigma$-dimer has the same dynamic properties as the Roper dimer and fulfills
an equation analogous to Eq.~\eqref{def-mass-and-width}. Then, one can approximate the width of
the $\sigma$ resonance $\Gamma_{\sigma}$ as 
\begin{align}
    \Gamma_{\sigma} \approx \left. \frac{1}{M_{\sigma}} \text{Im} \left \lbrace \Sigma_{\sigma} (p) \right \rbrace \right|_{p=M_{\sigma}} = \frac{h_2^2}{8 \pi M_{\sigma}^3} \lambda^{1/2} \left( M_{\sigma}^2, M_{\pi}^2, M_{\pi}^2 \right) \; . \label{decay-width-sigma}
\end{align}
Using phenomenological values for the mass and width of the $\sigma$-resonance, one can then fix the coupling
$h_2$. This method is more speculative, because we introduced the dimer as a constant field
and not as a dynamical one. Nonetheless, we do not
abandon this method yet, using it as an additional cross-check.

Our analysis of the $\sigma$-dimer can be repeated analogously for the $\Delta$-dimer. First, we take
the dressed propagator in Eq.~\eqref{dressed-delta-prop} and expand it like before to obtain  
\begin{align}
D_{\Delta} (p) = - \frac{\alpha_{\Delta}^{} m_{\Delta}^2 p^4 - i b p^2 \lambda^{1/2} \left( p^2, m_{N}^2 , M_{\pi}^2
\right) }{m_{\Delta}^4 p^4 + b^2 \lambda \left( p^2, m_{N}^2 , M_{\pi}^2 \right)} \; , \quad b = \frac{g_2^2}{16 \pi}
\; , 
\end{align}
where we again used that $\alpha_{\Delta}^2=1$. The two different masses inside the K{\"a}ll{\'e}n function
give the propagator a more complex structure. After some algebra the denominator can be rewritten as 
\begin{align}
\begin{split}
m_{\Delta}^4 p^4 + b^2 \lambda \left( p^2, m_{N}^2 , M_{\pi}^2 \right) &= m_{\Delta}^4 p^4 + b^2
\left( p^4 - 2 p^2 ( m_{N}^2 + M_{\pi}^2 ) + (m_{N}^2 - M_{\pi}^2)^2 \right) \\
&= \left( m_{\Delta}^4 + b^2 \right) \left( p^2 - \mu_{\Delta}^2 + i \nu \right) \left( p^2 - \mu_{\Delta}^2
- i \nu \right) \; ,
\end{split}
\end{align}
with 
\begin{align}
\mu_{\Delta}^2 = \frac{b^2 ( m_{N}^2 + M_{\pi}^2 )}{m_{\Delta}^4 + b^2} \; , \quad \text{and} \quad
\nu = \frac{b}{m_{\Delta}^4 + b^2} \sqrt{ m_{\Delta}^4 \left( m_{N}^2 - M_{\pi}^2 \right)^2 - 4 b^2 m_{N}^2 M_{\pi}^2 } \; . 
\end{align}
In the case of two equal masses in the K{\"a}ll{\'e}n function, the result from the $\sigma$-dimer can
be restored. All together, we have 
\begin{align}
D_{\Delta} (p) = - \frac{\alpha_{\Delta}^{} m_{\Delta}^2}{m_{\Delta}^4 + b^2} \, \frac{p^4}{\left( p^2 - \mu_{\Delta}^2 \right)^2 + \nu^2} + \frac{i b}{m_{\Delta}^4 + b^2} \, \frac{p^2 \lambda^{1/2} \left( p^2, m_{N}^2 , M_{\pi}^2 \right) }{\left( p^2 - \mu_{\Delta}^2 \right)^2 + \nu^2} \; . 
\end{align}
One observes that the propagator does not have poles on the real axis, in contrast to the $\sigma$ case.
An imaginary part emerges above the pion-nucleon threshold, $p^2 = s > (m_{N} + M_{\pi})^2$, and the relation to
the $\pi N$-scattering length $a_{\pi N}$ reads
\begin{align}
\left| \vec{q} \right| \cot \delta_{\pi N}  \left( s \right)
= \left| \vec{q} \right| \frac{ \text{Re} \left\lbrace T_{\pi N \to \pi N} (s) \right\rbrace }{\text{Im}
\left\lbrace T_{\pi N \to \pi N} (s) \right\rbrace} = \left| \vec{q} \right| \frac{ \text{Re}
\left\lbrace D_{\Delta} \left( s \right) \right\rbrace }{\text{Im} \left\lbrace D_{\Delta} \left( s \right)
\right\rbrace} = + \frac{1}{a_{\pi N}} + \mathcal{O} \big( \left| \vec{q} \right|^2 \big) \; .    
\end{align}
The pion-nucleon phase shift is denoted by $\delta_{\pi N}  \left( s \right)$ and $T_{\pi N \to\pi N} (s)\propto
D_{\Delta} (s)$ is the pion-nucleon scattering amplitude. With $\lambda^{1/2} \left( s, m_{N}^2 , M_{\pi}^2 \right) =
2 \sqrt{s} \left| \vec{q} \right|$ and $\sqrt{s} = \sqrt{m_{N}^2 + \left| \vec{q} \right|^2} + \sqrt{M_{\pi}^2
  + \left| \vec{q} \right|^2}$, we find  
\begin{align}
a_{\pi N} M_{\pi} = - \frac{2 b M_{\pi}}{\alpha_{\Delta}^{} m_{\Delta}^2 (m_{N} + M_{\pi})} = - \frac{g_2^2 M_{\pi}}{8 \pi \alpha_{\Delta}^{} m_{\Delta}^2 (m_{N} + M_{\pi})} \; . \label{scattering-length-delta-dimer}
\end{align}
The experimental value of the scattering length in the isospin $I = 3/2$ channel from the Roy-Steiner analysis
is $a_{N \pi}^{I = 3/2} M_{\pi}^{} = (-86.3 \pm 1.8) \times 10^{-3}$~\cite{Hoferichter:2015hva}, which
fixes the value of $\alpha_{\Delta}$ to be $+1$. Analogously to the $\sigma$-case, one can also use
the decay width to deduce the coupling $g_2$. We then have 
\begin{align}
    \Gamma_{\Delta} \approx \left. \frac{1}{m_{\Delta}} \text{Im} \left \lbrace \Sigma_{\Delta} (p) \right \rbrace \right|_{p=m_{\Delta}} = \frac{g_2^2}{16 \pi m_{\Delta}^3} \lambda^{1/2} \left( m_{\Delta}^2, m_{N}^2, M_{\pi}^2 \right) \; , \label{decay-width-delta}
\end{align}
where we again stress that the above method of determining the coupling might be more speculative than using the
scattering length. The insights from this section will help us to determine the dimer contributions to the
Roper resonance self-energy. The numerical calculation of the dimer LECs will be discussed later in
section~\ref{sec:Numerics}.

\section{Roper self-energy with dynamical dimer fields}
\label{sec:se-dimer}

Let us now come back to the self-energy contributions of the Roper resonance. From the $N\sigma$ channel, we obtain the loop-integral 
\begin{align}
\Sigma_{N \sigma} \left( p \right) = f_4^2 \int \frac{d^4 k}{(2 \pi)^4 i } S_{N} \left( p-k \right) D_{\sigma}
\left( k \right) \; .
\end{align}
In section~\ref{sec:se-roper} we already discussed that a constant dimer propagator $D_{\sigma}^0$ leads to a
vanishing integral. Therefore, we now consider the dressed propagator $D_{\sigma} \left( k \right)$
from Eq.~\eqref{dressed-sigma-prop} and obtain 
\begin{align}
\begin{split}
\Sigma_{N \sigma} \left( p \right) &= - f_4^2 \int \frac{d^4 k}{(2 \pi)^4 i } \frac{1}{2 \omega_{N} (\vec{p} - \vec{k}) \big[ \omega_{N} (\vec{p} - \vec{k}) - (p_0 - k_0) - i \epsilon \big]} \, \frac{1}{\alpha_{\sigma}^{} M_{\sigma}^{2} + \Sigma_{\sigma}^{} (k) } \\
&= - f_4^2 \int \frac{d^4 k}{(2 \pi)^4 i } \frac{1}{2 \omega_{N} (\vec{p} - \vec{k}) \big[ \omega_{N} (\vec{p} - \vec{k}) - (p_0 - k_0) - i \epsilon \big]} \Bigg\lbrace \alpha_{\sigma}^{} M_{\sigma}^{2} + 2 h_2^2  \\ 
&\phantom{=} \; \times \int \frac{d^4 l}{(2 \pi)^4 i} \frac{1}{4 \omega_{\pi} (\vec{k} - \vec{l} \, ) \omega_{\pi} (\vec{l} \, ) \big[ \omega_{\pi} (\vec{k} - \vec{l} \, ) - (k_0 - l_0) - i \epsilon \big] \big[ \omega_{\pi} (\vec{l} \, ) - l_0 - i \epsilon \big] }  \Bigg\rbrace^{-1} , 
\end{split}
\end{align}
where we have used the $\sigma$-dimer self-energy from Eq.~\eqref{sigma-self-energy}. We can see that
the $l_0$ integration inside the $\sigma$ self-energy can be carried out right away according
to our findings in section~\ref{sec:se-roper}. We then arrive at 
\begin{align}
\begin{split}
\Sigma_{N \sigma} \left( p \right) &= - \frac{f_4^2}{\alpha_{\sigma}^{} M_{\sigma}^{2}} \int \frac{d^4 k}{(2 \pi)^4 i } \frac{1}{2 \omega_{N} (\vec{p} - \vec{k}) \big[ \omega_{N} (\vec{p} - \vec{k}) - (p_0 - k_0) - i \epsilon \big]} \\ 
&\phantom{=} \; \times \Bigg\lbrace 1 +  \frac{2 h_2^2}{\alpha_{\sigma}^{} M_{\sigma}^{2}} \int \frac{d^3 l}{(2 \pi)^3} \frac{1}{4 \omega_{\pi} (\vec{k} - \vec{l} \, ) \omega_{\pi} (\vec{l} \, ) \big[ \omega_{\pi} (\vec{k} - \vec{l} \, ) + \omega_{\pi} (\vec{l} \, ) - k_0 - i \epsilon \big] }  \Bigg\rbrace^{-1} . 
\end{split}
\end{align}
The next step is to integrate out the remaining time component $k_0$, which is a bit more challenging.
For this, we use again Cauchy's theorem, going first to the rest-frame of the $N \sigma$-system,
i.e. $p = (E , \vec{0})$. We expand then the propagator of the $\sigma$-dimer into a geometric series 
\begin{align}
\begin{split}
\Sigma_{N \sigma} \left( E \right) &= - \frac{f_4^2}{\alpha_{\sigma}^{} M_{\sigma}^{2}} \int \frac{d^4 k}{(2 \pi)^4 i } \frac{1}{2 \omega_{N} (\vec{k}) \big[ k_0 - ( E - \omega_{N} (\vec{k}) + i \epsilon ) \big]} \\ 
&\phantom{=} \; \times \Bigg\lbrace 1 -  \frac{2 h_2^2}{\alpha_{\sigma}^{} M_{\sigma}^{2}} \int \frac{d^3 l}{(2 \pi)^3} \frac{1}{4 \omega_{\pi} (\vec{k} - \vec{l} \, ) \omega_{\pi} (\vec{l} \, ) \big[ k_0 - ( \omega_{\pi} (\vec{k} - \vec{l} \, ) + \omega_{\pi} ( \vec{l} \, ) - i \epsilon ) \big] }  \Bigg\rbrace^{-1} \\
&= - \frac{f_4^2}{\alpha_{\sigma}^{} M_{\sigma}^{2}} \int \frac{d^4 k}{(2 \pi)^4 i } \frac{1}{2 \omega_{N} (\vec{k}) \big[ k_0 - ( E - \omega_{N} (\vec{k}) + i \epsilon ) \big]} \\ 
&\phantom{=} \; \times \Bigg\lbrace 1 +   \frac{2 h_2^2}{\alpha_{\sigma}^{} M_{\sigma}^{2}} \int \frac{d^3 l}{(2 \pi)^3} \frac{1}{4 \omega_{\pi} (\vec{k} - \vec{l} \, ) \omega_{\pi} (\vec{l} \, ) \big[ k_0 - ( \omega_{\pi} (\vec{k} - \vec{l} \, ) + \omega_{\pi} ( \vec{l} \, ) - i \epsilon ) \big] } \\
&\phantom{= = =} \; \, + \left( \frac{2 h_2^2}{\alpha_{\sigma}^{} M_{\sigma}^{2}} \right)^2 \left[ \int \frac{d^3 l}{(2 \pi)^3} \frac{1}{4 \omega_{\pi} (\vec{k} - \vec{l} \, ) \omega_{\pi} (\vec{l} \, ) \big[ k_0 - ( \omega_{\pi} (\vec{k} - \vec{l} \, ) + \omega_{\pi} ( \vec{l} \, ) - i \epsilon ) \big] } \right]^2  \\
&\phantom{= = =} \; \, + \left( \frac{2 h_2^2}{\alpha_{\sigma}^{} M_{\sigma}^{2}} \right)^3 \left[ \int \frac{d^3 l}{(2 \pi)^3} \frac{1}{4 \omega_{\pi} (\vec{k} - \vec{l} \, ) \omega_{\pi} (\vec{l} \, ) \big[ k_0 - ( \omega_{\pi} (\vec{k} - \vec{l} \, ) + \omega_{\pi} ( \vec{l} \, ) - i \epsilon ) \big] } \right]^3 \\
&\phantom{= = =} \; \, + \ldots \Bigg\rbrace \; . 
\end{split}
\end{align}
Note that we rewrote the denominators containing the $k_0$ integration variable to better exhibit the
pole structure of the expression. The nucleon propagator has a pole in the upper complex plane
($k_0\in\mathds{C}$), whereas all propagators appearing in the geometric series have their pole in the
lower plane. We choose the pole of the nucleon propagator and close the contour around the upper half of
the complex plane. The first appearing $k_0$-integral is the already discussed tadpole diagram
\begin{align}
\mathcal{I}_{0} = \int_{-\infty}^{+\infty} \frac{d k_0}{2 \pi i } \frac{1}{\big[ k_0 - ( E - \omega_{N} (\vec{k})
+ i \epsilon ) \big]}  \; , 
\end{align}
which we replace with its residue in the upper complex plane, i.e. $\mathcal{I}_{0} = 1$, according to our arguments from
section~\ref{sec:se-roper}. The next integrals can be summarized by the following expression 
\begin{align}
\begin{split}
    \mathcal{I}_{n} &= \int_{-\infty}^{+\infty} \frac{d k_0}{2 \pi i } \frac{1}{\big[ k_0 - ( E - \omega_{N} (\vec{k}) + i \epsilon ) \big]} \\
    &\phantom{= \int_{-\infty}^{+\infty} \frac{d k_0}{2 \pi i }} \; \times \left[ \int \frac{d^3 l}{(2 \pi)^3} \frac{1}{4 \omega_{\pi} (\vec{k} - \vec{l} \, ) \omega_{\pi} (\vec{l} \, ) \big[ k_0 - ( \omega_{\pi} (\vec{k} - \vec{l} \, ) + \omega_{\pi} ( \vec{l} \, ) - i \epsilon ) \big] } \right]^n  \; , 
\end{split}
\end{align}
where $n$ is a positive integer fulfilling $n \geq 1$. For $n=1$ we obtain a similar $k_0$-integral as in
$J_{N \pi}$ from Eq.~\eqref{J-pi-N}, which can be evaluated analogously. Choosing the contour around the
upper pole we obtain 
\begin{align}
    \mathcal{I}_{1} =   \int \frac{d^3 l}{(2 \pi)^3} \frac{1}{4 \omega_{\pi} (\vec{k} - \vec{l} \, ) \omega_{\pi} (\vec{l} \, ) \big[ E - \omega_{N} (\vec{k}) - \omega_{\pi} (\vec{k} - \vec{l} \, ) - \omega_{\pi} ( \vec{l} \, ) + i \epsilon  \big] }   \; .  
\end{align}
If $n>1$, the integral looks more complicated, however, there is still just one pole in the upper complex
plane resulting in a single residue. We can therefore deduce that 
\begin{align}
    \mathcal{I}_{n} =  \left[ \int \frac{d^3 l}{(2 \pi)^3} \frac{1}{4 \omega_{\pi} (\vec{k} - \vec{l} \, ) \omega_{\pi} (\vec{l} \, ) \big[ E - \omega_{N} (\vec{k}) - \omega_{\pi} (\vec{k} - \vec{l} \, ) - \omega_{\pi} ( \vec{l} \, ) + i \epsilon  \big] } \right]^n   \; .  
\end{align}
Using these results, the self-energy is given by 
\begin{align}
\begin{split}
\Sigma_{N \sigma} \left( E \right) &= - \frac{f_4^2}{\alpha_{\sigma}^{} M_{\sigma}^{2}} \int \frac{d^3 k}{(2 \pi)^3 } \frac{1}{2 \omega_{N} (\vec{k})}  \\
&\phantom{=} \times \Bigg\lbrace 1 + \frac{2 h_2^2}{\alpha_{\sigma}^{} M_{\sigma}^{2}} \int \frac{d^3 l}{(2 \pi)^3} \frac{1}{4 \omega_{\pi} (\vec{k} - \vec{l} \, ) \omega_{\pi} (\vec{l} \, ) \big[ E - \omega_{N} (\vec{k}) - \omega_{\pi} (\vec{k} - \vec{l} \, ) - \omega_{\pi} ( \vec{l} \, ) + i \epsilon  \big] } \\ 
&\phantom{=} + \left( \frac{2 h_2^2}{\alpha_{\sigma}^{} M_{\sigma}^{2}} \right)^2  \left[ \int \frac{d^3 l}{(2 \pi)^3} \frac{1}{4 \omega_{\pi} (\vec{k} - \vec{l} \, ) \omega_{\pi} (\vec{l} \, ) \big[ E - \omega_{N} (\vec{k}) -  \omega_{\pi} (\vec{k} - \vec{l} \, ) - \omega_{\pi} ( \vec{l} \, ) + i \epsilon \big] } \right]^2 \\
&\phantom{=} + \ldots \Bigg\rbrace \; ,  
\end{split}
\end{align}
which is again a geometric series that can be summed up to 
\begin{align}
\begin{split}
\Sigma_{N \sigma} \left( E \right) &= - f_4^2 \int \frac{d^3 k}{(2 \pi)^3 } \frac{1}{2 \omega_{N} (\vec{k})} \Bigg\lbrace \alpha_{\sigma}^{} M_{\sigma}^{2} \\ 
&\phantom{=} \; + 2 h_2^2 \int \frac{d^3 l}{(2 \pi)^3} \frac{1}{4 \omega_{\pi} (\vec{k} - \vec{l} \, ) \omega_{\pi} (\vec{l} \, ) \big[ \omega_{N} (\vec{k}) + \omega_{\pi} (\vec{k} - \vec{l} \, ) + \omega_{\pi} ( \vec{l} \, ) - E - i \epsilon  \big] } \Bigg\rbrace^{-1} . \label{N-sigma-self-energy-1}
\end{split}
\end{align}
This remaining expression for the $N \sigma$ self-energy now contains only the spatial integration over an
internal loop momentum $\vec{l}$ and an external momentum $\vec{k}$, which is a useful starting point for
a numerical evaluation. The integral in the denominator of the latter equation produces poles, when
the rest-frame energy $E$ equals the energy of a free nucleon and two pions 
\begin{align}
    E = \omega_{N} (\vec{k}) + \omega_{\pi} (\vec{k} - \vec{l} \, ) + \omega_{\pi} ( \vec{l} \, ) \;.
    \label{eq:3b-onshellconfig}
\end{align}
In other words, we encounter exactly the three particle on-shell configuration $N\pi\pi$ that is crucial to
describe the dynamics of the Roper system. We can analyze the result in Eq.~\eqref{N-sigma-self-energy-1}
a little further and see what happens, when the $\sigma$-dimer becomes stable. In this case, we assume
that $h_2 \to 0$, which leads to a vanishing integral over the internal momentum $\vec{l}$, so that
the dimer propagator becomes constant, i.e. 
\begin{align}
\begin{split}
\Sigma_{N \sigma} \left( E \right) &= - \frac{f_4^2}{\alpha_{\sigma}^{} M_{\sigma}^{2}} \int \frac{d^3 k}{(2 \pi)^3 } \frac{1}{2 \omega_{N} (\vec{k})} = - \frac{f_4^2}{2 \alpha_{\sigma}^{} M_{\sigma}^{2}} \int \frac{d^3 k}{(2 \pi)^3 } \frac{1}{ \sqrt{ |\vec{k}|^2 + m_N^2 } } \; , 
\end{split}
\end{align}
which is a regular integral and vanishes in dimensional regularization. This we have already observed in
section~\ref{sec:se-roper} and, hence, agrees with our expectation.

A similar calculation can also be performed for the $\Delta$-dimer case. Its self-energy contribution
to the Roper with the dressed dimer propagator is given by 
\begin{align}
\begin{split}
\Sigma_{\Delta \pi} \left( p \right) &= f_3^2 \int \frac{d^4 k}{(2 \pi)^4 i } S_{\pi} \left( p-k \right) D_{\Delta} \left( k \right) \\
&= - f_3^2 \int \frac{d^4 k}{(2 \pi)^4 i } \frac{1}{2 \omega_{\pi} (\vec{p} - \vec{k}) \big[ \omega_{\pi} (\vec{p} - \vec{k}) - (p_0 - k_0) - i \epsilon \big]} \, \frac{1}{ \alpha_{\Delta}^{} m_{\Delta}^{2} + \Sigma_{\Delta}^{} (k) } \; , 
\end{split}
\end{align}
and after integrating out the $k_0$ component we arrive at 
\begin{align}
\begin{split}
\Sigma_{\Delta \pi} \left( E \right) &= - f_3^2 \int \frac{d^3 k}{(2 \pi)^3 } \frac{1}{2 \omega_{\pi} (\vec{k})} \Bigg\lbrace \alpha_{\Delta}^{} m_{\Delta}^{2} \\
&\phantom{=} \; + g_2^2 \int \frac{d^3 l}{(2 \pi)^3} \frac{1}{4 \omega_{\pi} (\vec{k} - \vec{l} \, ) \omega_{N} (\vec{l} \, ) \big[ \omega_{\pi} (\vec{k}) + \omega_{\pi} (\vec{k} - \vec{l} \, ) + \omega_{N} ( \vec{l} \, ) - E - i \epsilon  \big] } \Bigg\rbrace^{-1} . \label{Delta-pi-self-energy-1}
\end{split}
\end{align}
This result looks similar to Eq.~\eqref{N-sigma-self-energy-1}, only the LECs differ. Both dimer field
self-energy contributions to the Roper resonance mass will be investigated next. From here on, however, we will
work in a finite volume, which is explored in the next section.

\section{Finite-volume formalism}
\label{sec:FVformalism}

In this section, we consider the Roper resonance in a finite volume (FV) and introduce the corresponding
formalism. Since lattice QCD calculations are performed on a space-time lattice of finite size, the
system under investigation is always confined in a finite volume, which limits its spacial (and time)
extent. The finite volume influences the particle system and leads to so-called finite-volume effects.
We now place the Roper resonance system in a cubic box of length $L$ and calculate the finite-volume energy eigenvalues (in the following referred to as `energy levels').
This allows us to compare the energy levels from our effective approach with
lattice QCD spectra of the Roper. Note that for simplicity we keep the time direction continuous.

In a finite volume the loop integral of the spatial momenta is replaced by an infinite,
three-dimensional sum while the integration over the time component remains unchanged  
\begin{align}
\int \frac{d^3 k}{(2 \pi)^3} \left(\ldots \right) \mapsto \frac{1}{L^3} \sum_{\vec{k}}
\left(\ldots \right) \qquad\text{for}\quad
\vec{k} =  \frac{2 \pi}{L} \vec{n} \; , \quad \vec{n} \in \mathbb{Z}^3 \; . \label{discretemomenta}
\end{align}
These changes naturally influence the self-energy of the Roper resonance as well. In particular,
the poles of the FV Roper-propagator arise when
\begin{align}
2 \omega_{R} (\vec{p}) \left[ \omega_{R} (\vec{p}) - p_0  \right] - \Sigma_{R}^L (p_0, \vec{p}) = 0 \; , 
\label{FVpoles}
\end{align}
where $\Sigma_{R}^L (p_0, \vec{p})$ denotes the self-energy of the Roper in the finite box. Choosing again
the rest-frame, $p_0 = E$ and $\vec{p} = 0$, we can reformulate Eq.~\eqref{FVpoles} so that we obtain
an equation for the energy levels in the finite volume. We find 
\begin{align}
2 m_{R0} \left( m_{R0} - E  \right) = \Sigma_{R}^L (E) \quad \Leftrightarrow \quad m_{R0} - E
= \frac{1}{2 m_{R0}} \Sigma_{R}^L (E) \; , 
\end{align}
which is the master equation for the finite-volume energy levels of the Roper resonance in this framework.
A remaining problem is the appearance of the bare mass $m_{R0}$ in the equation. However, for the numerical
calculation of the energy levels we set the bare mass equal to the physical mass $m_R$. After this, one arrives at 
\begin{align}
   m_R - E - \frac{1}{2 m_{R}} \Sigma_{R}^L (E) = 0 \; , 
   \label{eq:Energy-levels-FV}
\end{align}
which is the equation we will work with. Note that this self-energy equation shares similarities with the usual
three-body quantization conditions~\cite{Hansen:2014eka,Hammer:2017kms,Mai:2017bge}, e.g. by accounting
for three-particle on-shell configurations, see Eq.~\eqref{eq:3b-onshellconfig}. 

Next, we have to determine the exact form of $\Sigma_{R}^L (E)$. As we have seen in Eq.~(\ref{Roper-Self-Energy}),
the Roper self-energy consists of three contributions, which is also true in the finite volume, 
\begin{align}
    \Sigma_{R}^L (E) = \Sigma_{N \pi}^L (E) + \Sigma_{N \sigma}^L (E) + \Sigma_{\Delta \pi}^L (E) \; . \label{Roper-Self-Energy-FV}
\end{align}
Let us start with $\Sigma_{N \pi}^L (E)$, which is given by 
\begin{align}
    \Sigma_{N \pi}^L (E) = f_2^2 J_{N \pi}^L (E)  \; , 
\end{align}
where $J_{N \pi}^L$ is the finite-volume version of integral $J_{N \pi}$ from Eq.~\eqref{J-pi-N}. We have seen
in the discussion of Eq.~\eqref{J-pi-N}, that the first step is integrating over the time component of
the momentum. One then arrives at Eq.~\eqref{J-pi-N-2} and the spatial integral is now replaced by a sum leading to 
\begin{align}
\begin{split}
J_{N \pi}^L \left( E \right) = \frac{1}{L^3} \sum_{\vec{k}} \, \frac{1}{4 \omega_{N} (\vec{k}) \omega_{\pi} (\vec{k})
\big[ \omega_{N}( \vec{k}) + \omega_{\pi} (\vec{k}) - E  \big] } \; , \label{J-pi-N-FV} 
\end{split}
\end{align}
in the rest-frame. We expand the integrand again according to Eq.~\eqref{integrand-rewritten} and get 
\begin{align}
J_{N \pi}^L \left( E \right) = \frac{1}{L^3} \sum_{\vec{k}} \, \frac{1}{2 E} \frac{1}{|\vec{k}|^2 - q^2 (E)} + \ldots \; , 
\end{align}
where the ellipses denote the  remaining regular terms. These terms, as we have observed, vanish in the
infinite volume and lead to contributions proportional to $\exp(-M_\pi L)$ in the finite volume.
The latter effects are sub-leading to the other effects discussed here and are neglected in what follows.
Thus, analogous to the infinite-volume case, also in the finite volume only the term containing
the pole survives. Using Eq.~\eqref{discretemomenta}, we can write 
\begin{align}
J_{N \pi}^L \left( E \right) = \frac{1}{8 \pi E L} \sum_{\vec{n}} \, \frac{1}{|\vec{n}|^2 - \tilde{q}^2 (E)}
= \frac{1}{4 \pi^{3/2} E L} \mathcal{Z}_{00} \left( 1, \tilde{q}^2 (E) \right)  \; , \label{J-pi-N-FV-result}
\end{align}
where we rescaled the variable $q(E)$ as $\tilde{q}^2 (E) = L^2 q^2(E) / (2 \pi)^2$ and used the standard
L{\"u}scher Zeta-function~\cite{Luscher:1990ux}.
The finite-volume expression for the $N\pi$ contribution is then given by 
\begin{align}
\Sigma_{N \pi}^L (E) = \frac{f_2^2}{4 \pi^{3/2} E L} \mathcal{Z}_{00} \left( 1, \tilde{q}^2 (E) \right) \; . 
\label{Sigma-N-pi-FV-correction}  
\end{align}

Next, we turn to the self-energy contribution with the nucleon and $\sigma$-dimer field, $\Sigma_{N \sigma}^L$.
For this, we take the  result from Eq.~\eqref{N-sigma-self-energy-1} and replace the integrals by sums 
\begin{align}
\begin{split}
\Sigma_{N \sigma}^L \left( E \right) &= - \frac{f_4^2}{L^3}  \sum_{\vec{k}} \frac{1}{2 \omega_{N} (\vec{k})} \Bigg\lbrace \alpha_{\sigma}^{} M_{\sigma}^{2} \\
&\phantom{=} \; + \frac{2 h_2^2}{L^3} \sum_{\vec{l}} \frac{1}{4 \omega_{\pi} (\vec{k} - \vec{l} \, ) \omega_{\pi} (\vec{l} \, ) \big[ \omega_{N} (\vec{k}) + \omega_{\pi} (\vec{k} - \vec{l} \, ) + \omega_{\pi} ( \vec{l} \, ) - E - i \epsilon  \big] } \Bigg\rbrace^{-1} . \label{N-sigma-self-energy-FV} 
\end{split}
\end{align}
Analogously, the finite-volume contribution with pion and $\Delta$-dimer field has the form 
\begin{align}
\begin{split}
\Sigma_{\Delta \pi}^L \left( E \right) &= - \frac{f_3^2}{L^3}  \sum_{\vec{k}} \frac{1}{2 \omega_{\pi} (\vec{k})} \Bigg\lbrace \alpha_{\Delta}^{} m_{\Delta}^{2} \\
&\phantom{=} \; + \frac{g_2^2}{L^3} \sum_{\vec{l}} \frac{1}{4 \omega_{\pi} (\vec{k} - \vec{l} \, ) \omega_{N} (\vec{l} \, ) \big[ \omega_{\pi} (\vec{k}) + \omega_{\pi} (\vec{k} - \vec{l} \, ) + \omega_{N} ( \vec{l} \, ) - E - i \epsilon  \big] } \Bigg\rbrace^{-1} . \label{Delta-pi-self-energy-FV} 
\end{split}
\end{align}
These two expressions can readily be worked out numerically, however, a cutoff is naturally required to
tame the otherwise infinite sums.  In our calculations, the outer sum runs to $ L |\vec{k}| / (2 \pi)
\approx 3$ to ensure a similar energy coverage as in Ref.~\cite{Severt:2020jzc}. 
The inner momentum is carried out until $ L |\vec{l}| / (2 \pi) \approx 10$, so that $|\vec{l}| >
|\vec{k}|$ is fulfilled. With these results we can now calculate the energy levels of the
Roper resonance numerically.

\section{Numerical calculation}
\label{sec:Numerics}

The energy spectrum of the Roper resonance system is determined by numerically finding solutions of
\begin{align}
m_R - E = \frac{1}{2 m_R} \left( \Sigma_{N \pi}^L (E) + \Sigma_{N \sigma}^L (E) + \Sigma_{\Delta \pi}^L (E) \right) \; ,
\label{eq:Energy-levels-FV-final} 
\end{align}
with respect to $E\in\mathds{R}$. Here $\Sigma_{N \pi}^L$, $\Sigma_{N \sigma}^L$ and $\Sigma_{\Delta \pi}^L (E)$
are given in Eqs.~\eqref{Sigma-N-pi-FV-correction},~\eqref{N-sigma-self-energy-FV}
and~\eqref{Delta-pi-self-energy-FV}, respectively.
Note that during the derivation of Eq.~\eqref{eq:Energy-levels-FV-final}, we have seen that certain
contributions decrease exponentially for large $L$, which we already neglected. We therefore have to
choose $L$ large enough to justify these approximations. An avoided level crossing in the energy
spectrum is expected around the Roper resonance mass.

For the hadron masses we use the numerical values from Ref.~\cite{Gegelia:2016xcw} and the
PDG~\cite{ParticleDataGroup:2022pth}. Specifically, the Roper resonance mass is $m_R = 1365\,$MeV,
the pion mass in the isospin-limit is set to $M_{\pi} = 139\,$MeV and the nucleon mass is $m_N = 939\,$MeV.  
To fix the LECs $\{f_2,f_3,f_4,g_2,h_2\}$, we need further observables. The self-energy $\Sigma_{N \pi}$, for
example, is proportional to the LEC $f_2$, see Eq.~\eqref{self-energy-pi-N-2}. This constant is connected
to the decay of the Roper resonance into a nucleon and a pion. According to the
PDG~\cite{ParticleDataGroup:2022pth} the width of the Roper is $\Gamma_{R} = 190\,$MeV, where the
decay into a nucleon and a pion contributes to (approximately) $65\%$, i.e. $\Gamma_{R \rightarrow N \pi} = 123.5\,$MeV.
The other $35\%$ contribute to the decay with two pions in the final state, $\Gamma_{R \rightarrow N \pi \pi}
= 66.5\,$MeV. However, this final state can be reached by the different intermediate $N \sigma$ or $\Delta \pi$
states. The decay widths into these unstable intermediate states are approximately $\Gamma_{R \rightarrow N \sigma}
= 38\,$MeV and $\Gamma_{R \rightarrow \Delta \pi} = 28.5\,$MeV~\cite{ParticleDataGroup:2022pth}. We can use these decay widths to fit some of the
LECs, like $f_2$. From Eq.~\eqref{def-mass-and-width}, we know that the width is connected to the
imaginary part of the self-energy. We find 
\begin{align}
\Gamma_{R \rightarrow N \pi} \approx \left. \frac{1}{m_{R}} \text{Im} \left \lbrace \Sigma_{N \pi} (E) \right \rbrace
\right|_{E=m_R} = \frac{1}{m_{R}} \text{Im} \left \lbrace \Sigma_{N \pi} (m_R) \right \rbrace \; ,  
\end{align}
where $\Sigma_{N \pi} (m_R)$ consists solely of known parameters, except $f_2$. Using the PDG estimate for
$\Gamma_{R \rightarrow N \pi}$, we find 
\begin{align}
   \Gamma_{R \rightarrow N \pi} = 7.24 \times 10^{-3} f_2^2 \, \text{GeV}^{-1} \; \Leftrightarrow \; f_2^{} = \pm 4.13 \, \text{GeV} \; .    
\end{align}
The sign of $f_2$ cannot be determined through this procedure, but this does not matter for our further analysis. 
The matter becomes more complicated when looking at the self-energy contributions including dimer fields.
The self-energy $\Sigma_{N \sigma}$, for example, contains three parameters $h_2$, $M_{\sigma}$ and $f_4$ that
have to be determined. We set $M_{\sigma}$ to the physical mass of the $f_0(500)$, since this scale appears
in the $\sigma$-dimer propagator. The PDG~\cite{ParticleDataGroup:2022pth} estimates for the $f_0(500)$ are
$M_{\sigma} = (400 - 550)\,$MeV and $\Gamma_{\sigma}= (400 - 700)\,$MeV. 
For simplicity we take the lower values, assuming $M_{\sigma} = 400\,$MeV, which also fulfills
$(M_{\sigma} + m_N) < m_R$, and $\Gamma_{\sigma}= 400\,$MeV. For the self-energy contribution from the
$\Delta$-dimer, $\Sigma_{\Delta \pi}$, the unknown LECs are $g_2$ and $f_3$, and we also set $m_{\Delta}$ to
the physical delta mass. The mass and width of the delta resonance have been more accurately determined,
and we set them here to $m_{\Delta} = 1210\,$MeV and $\Gamma_{\Delta}= 100\,$MeV.

Using these phenomenological values we determine the unknown constants as follows. We begin with an
estimate for the constants $f_3$ and $f_4$. Assuming that $\sigma$ and $\Delta$ are stable final states
with the same kinematic behaviour as the nucleons and pions, their self-energy contributions to the Roper
resonance mass are given by 
\begin{align}
\Sigma_{N \sigma}^{\text{stable}} (E) = \frac{i f_4^2}{16 \pi E^2} \lambda^{1/2} \left( E^2, m_N^2, M_{\sigma}^2 \right) \; ,
\quad \text{and} \quad \Sigma_{\Delta \pi}^{\text{stable}} (E) = \frac{i f_3^2}{16 \pi E^2} \lambda^{1/2}
\left( E^2, m_{\Delta}^2, M_{\pi}^2 \right) \; .
\end{align}
Taking $\Sigma_{N \sigma}^{\text{stable}}$, for example, we can approximate the decay width of Roper going to
a $N \sigma$ final state by 
\begin{align}
\Gamma_{R \rightarrow N \sigma} \approx \left. \frac{1}{m_{R}} \text{Im} \left \lbrace \Sigma_{N \sigma}^{\text{stable}} (E)
\right \rbrace \right|_{E=m_R} = \frac{ f_4^2}{16 \pi m_R^3} \lambda^{1/2} \left( m_R^2, m_N^2, M_{\sigma}^2 \right) \; . 
\end{align}
Using our values for the decay width and masses, one arrives at $f_4 = \pm 3.82\,$GeV. An analogous calculation
with $\Sigma_{\Delta \pi}^{\text{stable}}$ leads to $f_3 = \pm 4.55\,$GeV, meaning that within this approximation
$f_3$ and $f_4$ are of the same magnitude. In the future, one might also consider lattice QCD data to
determine the numerical values for these constants, but, for now, we use the above estimations.

Next, we consider the LECs $h_2$ and $g_2$. As already stated in section~\ref{sec:Dimerfield}, these
constants can be related to the two-particle scattering lengths. For $h_2$, we found the relation
given in Eq.~\eqref{scattering-length-dimer-constants}. Using the $\sigma$ mass, $\alpha_{\sigma} = -1$
for an attractive interaction and the value $a^{I=0} M_{\pi} = 0.222$ for the $\pi \pi$-scattering length,
we obtain 
\begin{align}
h_2^2 = 8 \pi M_{\sigma}^2 (a^{I=0} M_{\pi}) \quad \Rightarrow \quad h_2^{} = \pm 0.95\,\text{GeV} \; .  
\end{align}
Now, we take a look what happens if we use the decay width to fix $h_2$. With Eq.~\eqref{decay-width-sigma}
and the PDG~\cite{ParticleDataGroup:2022pth} data above, we find 
\begin{align}
  h_2^2 = \frac{8 \pi M_{\sigma}^3 \Gamma_{\sigma}^{} }{\lambda^{1/2} \left( M_{\sigma}^2, M_{\pi}^2, M_{\pi}^2 \right)}
  \quad \Rightarrow \quad h_2^{} = \pm 2.36\,\text{GeV} \; , 
\end{align}
which is interestingly of the same order of magnitude albeit around two times larger than the prediction
from the scattering length. As of the coupling $g_2$, we use the $\pi N$-scattering length, $a_{\pi N}^{I=3/2}
M_{\pi}^{} = - 0.086$ with the delta resonance mass, $\alpha_{\Delta} = +1$ and the help of
Eq.~\eqref{scattering-length-delta-dimer}. This yields 
\begin{align}
g_2^2 = - 8 \pi \alpha_{\Delta}^{} m_{\Delta}^2 (m_N^2 + M_{\pi}^2) a_{N \pi}^{I=3/2} \quad \Rightarrow \quad g_2^{}
= \pm 4.96\,\text{GeV} \; , 
\end{align}
whereas using Eq.~\eqref{decay-width-delta} and the above value for $\Gamma_{\Delta}$ leads  to 
\begin{align}
    g_2^2 = \frac{16 \pi m_{\Delta}^3 \Gamma_{\Delta}^{} }{\lambda^{1/2} \left( m_{\Delta}^2, m_{N}^2, M_{\pi}^2 \right)} \quad \Rightarrow \quad g_2^{} = \pm 4.22\,\text{GeV} \; . 
\end{align}
We see that in the $\Delta$ case both ways to fix the LEC $g_2$ lead to approximately the same value. This
might be related to the  fact that the delta resonance has a Breit-Wigner shape to very good accuracy. It is
good to see that the particle-dimer approach is consistent with this by giving $g_2$ almost equally
from the scattering length and the decay width. 

Before turning to the prediction of the Roper finite-volume spectrum, we try to test the quality of the dimer LECs determination presented above. For this we turn to the $\sigma$-dimer, and concentrate
solely on the two-particle $\pi \pi$ final state. In the finite volume, the $\sigma$-dimer propagator is
given by 
\begin{align}
D_{\sigma} \left( E \right) = - \frac{1}{\alpha_{\sigma}^{} M_{\sigma}^2 + \Sigma_{\sigma}^{L} (E) } \; , \quad \text{with} \quad \Sigma_{\sigma}^{L} (E) = \frac{2 h_2^2}{L^3} \sum_{\vec{k}} \, \frac{1}{4 \omega_{\pi} (\vec{k}) \omega_{\pi} (\vec{k}) \big[ 2 \omega_{\pi} (\vec{k}) - E  \big] } \; , \label{two-particle-test-eq}
\end{align}
where we again restricted ourselves to the rest-frame ($p_0 = E$, $\vec{p}=0$). The poles of the propagator in Eq.~\eqref{two-particle-test-eq} correspond to the interacting finite-volume energy levels of the $\pi \pi$ system, i.e.  
\begin{align}
f_{\sigma} (E) := 1 + \frac{1}{\alpha_{\sigma}^{} M_{\sigma}^2} \Sigma_{\sigma}^{L} (E) \overset{!}{=} 0 \; .
\label{sigma-dimer-eq:Energy-levels-FV}
\end{align}
Using this formula we can compare the energy levels from the particle-dimer picture with lattice QCD results.
Before going to this we wish to remark that the latter condition is related to the well established L\"uscher's method~\cite{Luscher:1986pf,Luscher:1990ux}. This can be seen by using a similar decomposition as shown in Eq.~\eqref{integrand-rewritten} of the integrand in~\eqref{two-particle-test-eq}. In this pilot study of the proposed formalism, we stay with the condition~\eqref{sigma-dimer-eq:Energy-levels-FV} leaving a more quantitative discussion to future studies. 

Lattice studies on the $\sigma$ resonance have already been performed, see e.g. Refs.~\cite{Fu:2013ffa,Briceno:2016mjc, Briceno:2017qmb,Guo:2018zss,Liu:2016cba}. Here we consider results of the combined $I=0,1,2$ finite-volume  analysis~\cite{Mai:2019pqr} of GWQCD lattice results~\cite{Guo:2018zss,Culver:2019qtx,Guo:2016zos} obtained at two values of pion mass.
%
%
%
%
For both cases the $\pi \pi$ scattering length $a^{I=0}$, the $\sigma$ mass $M_{\sigma}$
and the width $\Gamma_{\sigma}$ have been determined
\begin{align}
    \text{Set 1}:  \;
    \nonumber&M_{\pi} = 0.224 \, \text{GeV} \; , \; M_{\pi}L = 3.3  \; , \;\\ 
    &M_{\sigma} = 0.502 \, \text{GeV} \; , \; \Gamma_{\sigma} = 0.350 \, \text{GeV} \; , \; a^{I=0} M_{\pi} = 0.699 \; , \label{Set1} \\
    \text{Set 2}: \; 
    &M_{\pi} = 0.315 \, \text{GeV} \; , \; M_{\pi}L = 4.6  \; ,\hspace{7cm}\nonumber\\
    &M_{\sigma} = 0.591 \, \text{GeV} \; , \; \Gamma_{\sigma} = 0.218 \, \text{GeV} \; , \; a^{I=0} M_{\pi} = 1.901 \; .  \label{Set2}
\end{align}
We now take each data set and calculate the LEC $h_2$ from the scattering length $a^{I=0} M_{\pi}$ and
width $\Gamma_{\sigma}$. For Set $1$, we obtain $h_2 = 2.10\,$GeV using the scattering length and
Eq.~\eqref{scattering-length-dimer-constants}, and $h_2 = 3.13 \,$GeV using the width and
Eq.~\eqref{decay-width-sigma}. For Set $2$, the scattering length leads to $h_2 = 4.08\,$GeV,
while Eq.~\eqref{decay-width-sigma} cannot be used. This is because of the large pion mass $M_{\pi} = 0.315\,$GeV
preventing the decay of the $\sigma$ meson into two pions.
\begin{figure}[t]
	\begin{center}  
		\includegraphics[width=0.49\linewidth]{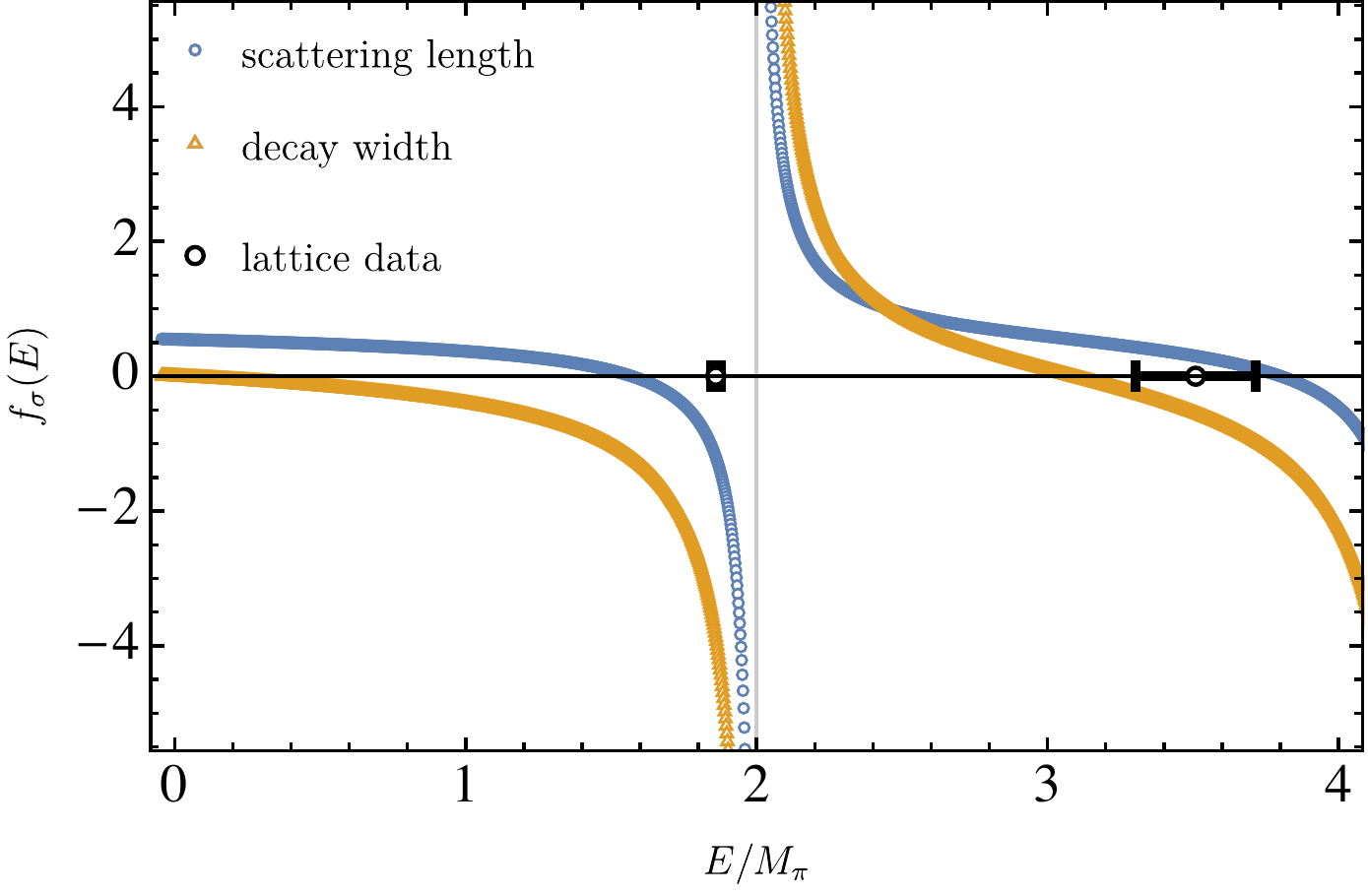}
		\hfill
		\includegraphics[width=0.49\linewidth]{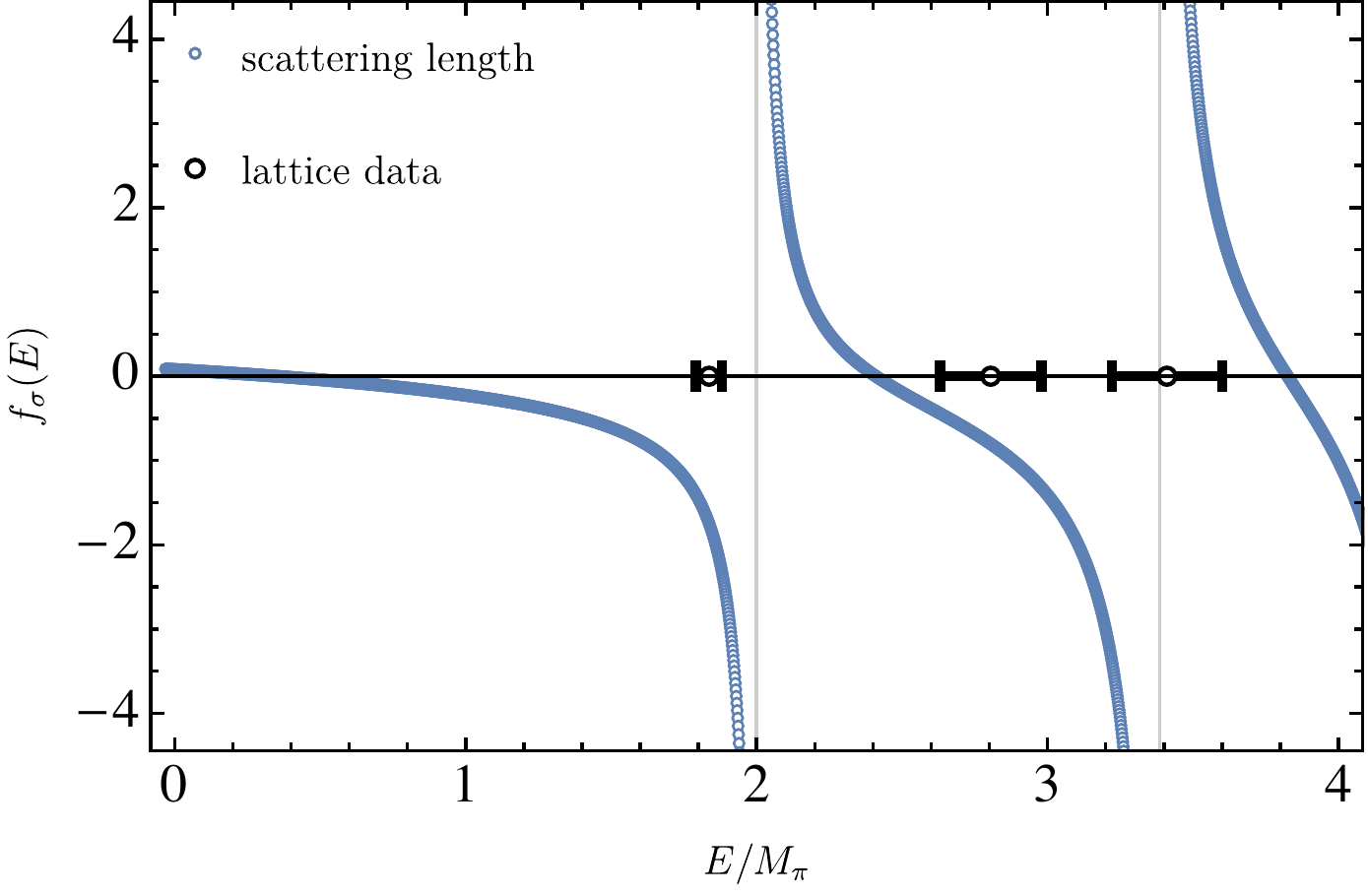}
	\end{center}
	\caption{
	  Predicted energy levels (zeroes of $f_\sigma(E)$) of the $\pi \pi$-system within the $\sigma$-dimer
          approach using Set $1$~\eqref{Set1} (left panel) and $2$~\eqref{Set2} (right panel). The blue curves
          show the function $f_{\sigma}(E)$ with $h_2$ determined from the two scattering lengths $a^{I=0} M_{\pi}$
          in each data set and the orange curve shows $f_{\sigma}(E)$ with $h_2$ determined from the decay width
          $\Gamma_{\sigma}$ (only for Set $1$). Black circles display the lattice results with errors
          from Ref.~\cite{Mai:2019pqr} and the grey vertical lines the non-interacting $\pi \pi$ energy eigenvalues.  
	} 
	\label{fig:Sigma-EnLevels-1}
\end{figure}
In principal, one could test the above procedure even further by using more lattice QCD data on the $\sigma$ meson for various pion masses from different working groups. However, this would go beyond the scope of this work and especially well beyond this qualitative check-up of the numerical estimation of the dimer LECs. A comparison of those data within our framework could be dedicated to future works.
The predicted two-body finite-volume spectrum for both data sets is depicted in Fig.~\ref{fig:Sigma-EnLevels-1}.
Therein, the left panel of Fig.~\ref{fig:Sigma-EnLevels-1} shows the function $f_{\sigma}(E)$ for data set~$1$ with
$h_2$ fixed by the scattering length (blue pionts) and by the decay width (orange pionts). The zeros of
this function show the energy levels for this two-pion system. The black circles are the lattice QCD results
from Ref.~\cite{Mai:2019pqr}. We observe that the levels from the blue curve lie very close to the lattice
results. The orange curve, on the other hand, still reproduces the first excited level above the
two-pion threshold at $\approx 1\sigma$, but the ground-state level is at odds with the lattice result. The zero for the ground state lies very close to $E/M_{\pi} \approx 0$. Since the driving term includes only momentum-independent structures we do not expect any predictive power from this formalism so far below threshold.
Therefore, the constant $h_2$ fixed by the scattering length leads to a better reproduction of the lattice
results. The right panel of Fig.~\ref{fig:Sigma-EnLevels-1} shows $f_{\sigma}(E)$ obtained with data set~$2$.
Here, as stated before, we only have the result from the scattering length estimation. The lattice results
are again depicted by the black circles. Overall, there is less agreement between the predicted levels and
those from the lattice. The ground-state level lies again well below $E/M_{\pi} = 1$ and merely the excited levels are somewhat close to the lattice QCD results. 
We emphasize again that the data from set~$2$ are determined by a pion mass much larger than set~$1$ and that
the $\sigma$-meson mass is smaller than two pion masses, which forbids the decay of $\sigma$ into two pions.
This is a condition that we did not take into account in our theoretical framework and it might explain the
large deviations between the dimer and lattice results.

There are two take-away messages from this analysis of the $\sigma$-dimer propagator and the corresponding
$\pi\pi$ finite-volume spectrum: First, we have seen that the particle-dimer approach works better for
smaller pion masses. This does not come as a surprise, since the dimer propagator is by construction a
constant at leading order. Second, we have seen that for lower pion mass the $\pi \pi $ scattering
length ensures a better description of the lattice QCD spectrum than the decay width of the $\sigma$ meson.
Hence, we will use the scattering length to fix the dimer LECs $h_2$ and $g_2$ for our calculation of the
Roper resonance energy levels. Finally, we note that no fit to the lattice data and, also, no similar study for the two-particle $N \pi$ scattering in the $\Delta$ channel (some lattice studies of $\Delta$-resonance can be found in Refs.~\cite{Meissner:2010ij,Alexandrou:2013ata,Alexandrou:2015hxa,Andersen:2017una,Alexandrou:2021plg,Silvi:2021uya}) are performed in this pioneering study.

\section{Results}
\label{sec:results}

\begin{figure}[t]
	\begin{center}  
		\includegraphics[width=0.6\linewidth]{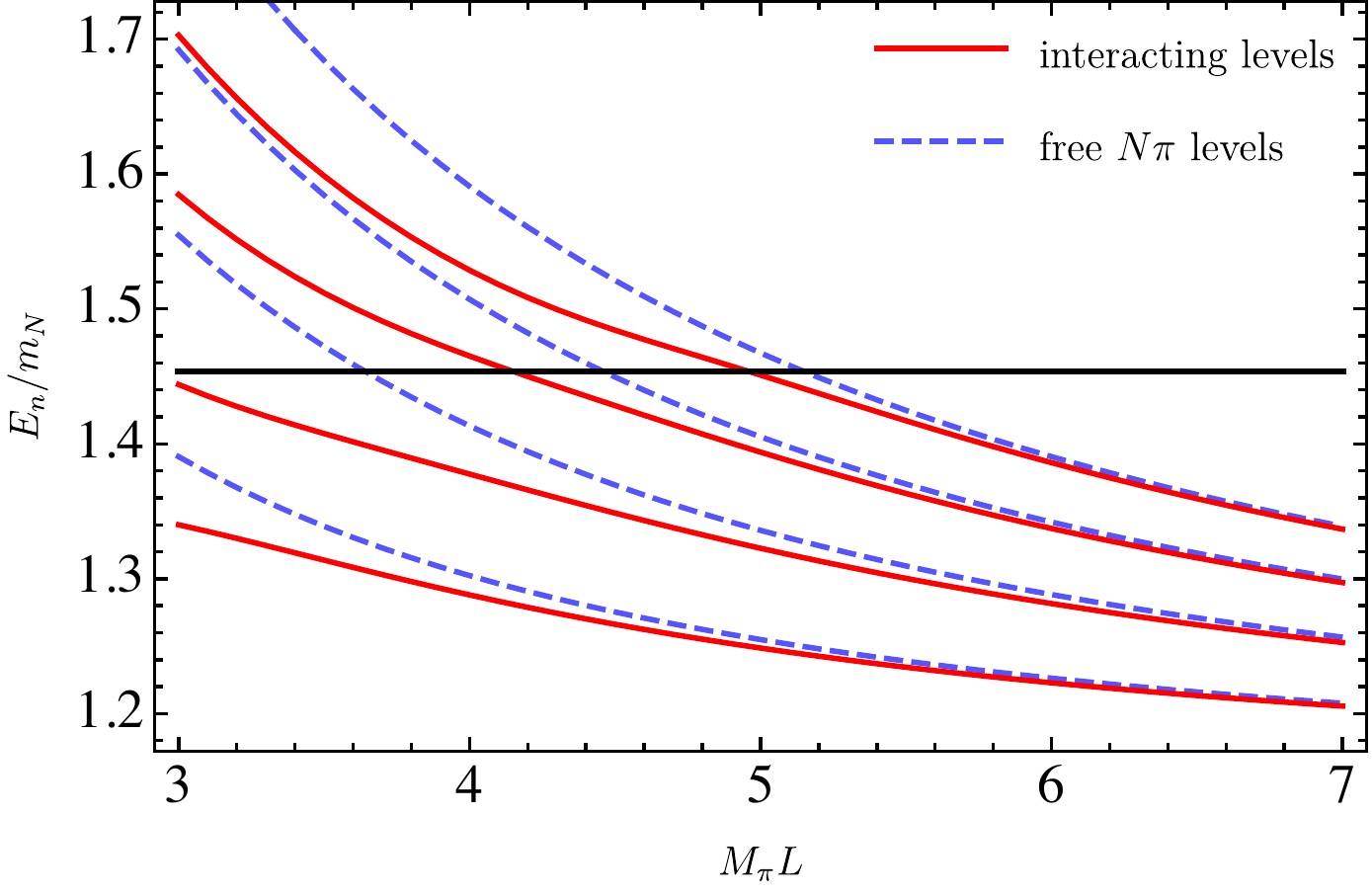}
	\end{center}
	\caption{
	  Energy levels for different box sizes $L$ considering only pion and nucleon as intermediate states.
          The red solid lines display the numerical results for the interacting energy levels and the blue
          dashed lines the free (non-interacting) energy levels of the pion and nucleon for
          $|\vec{n}_{1,2}|^2 = 1,2,3,4$ (lowest to highest curve). The thick solid black line marks
          the mass of the Roper resonance.
	} 
	\label{fig:FitNpi}
\end{figure}

Now that numerical values of constants are determined, we proceed with the determination of energy levels
of the Roper system for the three different channels $N\pi$, $N\sigma$ and $\Delta\pi$. After this, we also take a look at the coupled channel $N\pi/N\sigma$ and compare our obtained energy values with lattice QCD calculations. 
We note again that $\Delta$ and $\sigma$ fields are allowed to decay to $N\pi$ and $\pi\pi$ channels, respectively. Thus, these states can simply be seen as auxiliary degrees of freedom accounting for different configurations of the $N\pi\pi$ system.

\subsection{{\boldmath$N \pi$} channel}

First of all, we perform a numerical calculation including only the $\Sigma_{N \pi}$ contribution. That means
only pion and nucleon intermediate states are considered and we neglect the self-energy with the
$\sigma$-dimer and $\Delta$-dimer, i.e. we set $f_3=f_4=0$ for now. The obtained levels can be
compared with the results from Ref.~\cite{Severt:2020jzc}, which serves as a test for the 
theoretical framework. The results are displayed in Fig.~\ref{fig:FitNpi}, where the energy is given
in units of the nucleon mass $m_N$ and the box length $L$ is multiplied by the pion mass $M_{\pi}$ to obtain
a dimensionless quantity for the box size. 
The red solid lines denote the numerical results of $E$ for the respective energy levels while the blue
dashed lines denote the free energy levels of the pion-nucleon final states (also in units of $m_N$), i.e. 
\begin{align}
E_{\pi N}^{\text{free}} \left( \vec{n}_1 , \vec{n}_2 \right)
= \sqrt{m_N^2 + \left( \frac{2 \pi}{L} \right)^2 | \vec{n}_1 |^2 } + \sqrt{M_{\pi}^2
+ \left( \frac{2 \pi}{L} \right)^2 | \vec{n}_2 |^2 } \; .  \label{piNfree-levels}
\end{align} 
Here, $\vec{n}_1$ and $\vec{n}_2$ are the discretized momenta  of the nucleon and pion with $\vec{n}_1+\vec{n}_2=0$.
We restrict ourselves to the first four levels for simplicity. The thick solid black line corresponds to
the real part of the Roper resonance mass, i.e. $m_R/m_N \approx 1.45$, which is from here on called
the ``critical value''. We can see clear signs of avoided level crossing at small box sizes around
the critical value, i.e. the energy levels switch from one free energy level to another, most notably
between the free levels $|\vec{n}_{1,2}|=3$ and $|\vec{n}_{1,2}|=4$ in Fig.~\ref{fig:FitNpi}.
Overall, Fig.~\ref{fig:FitNpi} is in very good agreement with the result obtained in Ref.~\cite{Severt:2020jzc}
(for more comparisons, see  Ref.~\cite{Severt:2022eic}). This is a noteworthy result considering that the
present formalism is much simpler. In Ref.~\cite{Severt:2020jzc} the full Lagrangian from baryon chiral
perturbation theory has been used including Lorentz-, spin- and isospin-structure. Slight deviations in
the numerical results can be observed mostly for small values of $M_{\pi}L$ which is expected.
However, the general similarity between the numerical results is striking, making us optimistic
to proceed with this approach.

\subsection{{\boldmath$N \sigma$} channel}

\begin{figure}[t]
	\begin{center}  
		\includegraphics[width=0.7\linewidth]{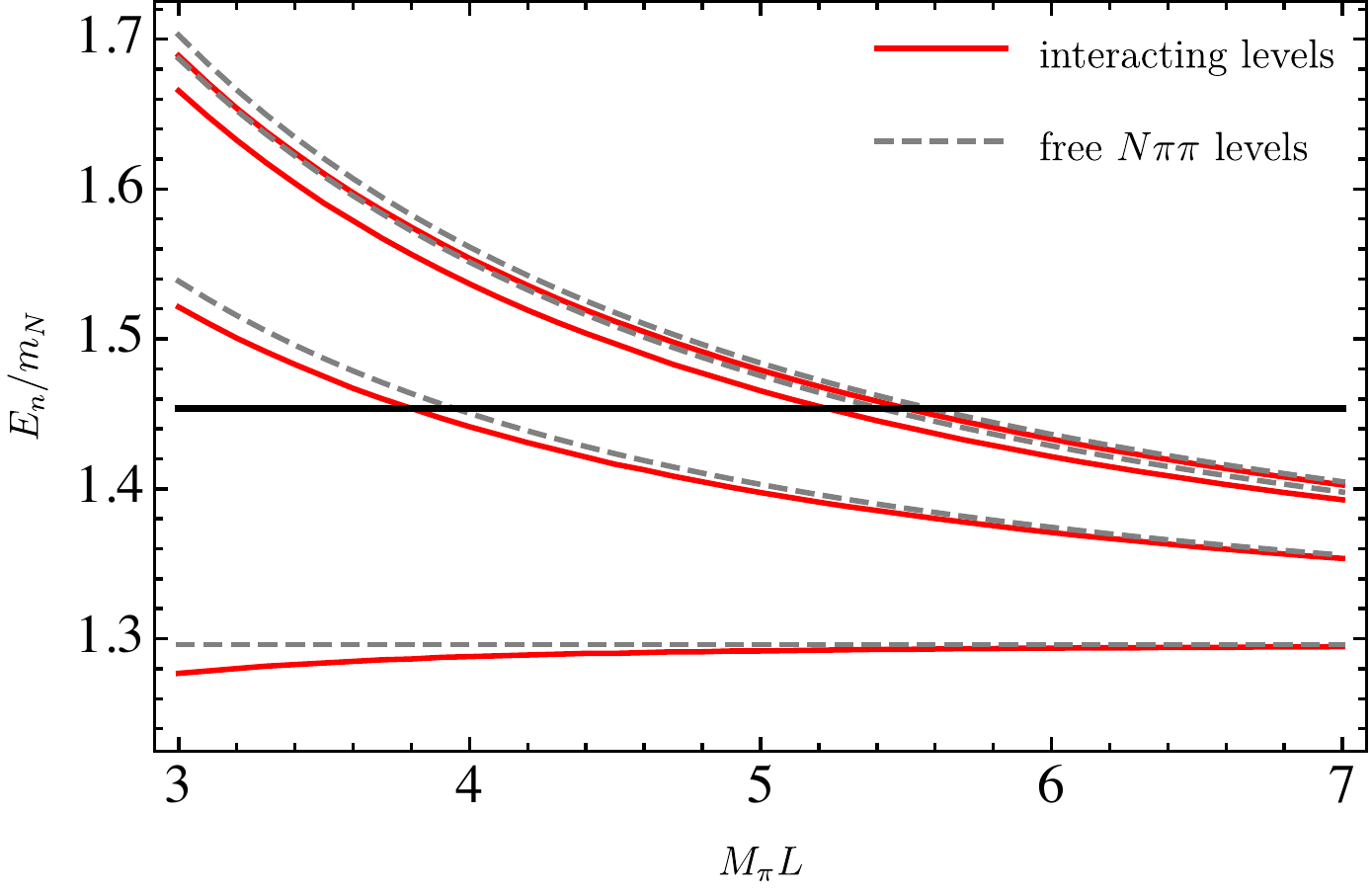}
	\end{center}
	\caption{
	  Roper energy levels for different box sizes $L$ considering only nucleon and $\sigma$-dimer as
          intermediate states. Red solid lines display the numerical results for the interacting energy
          levels and grey dashed lines the free (non-interacting) lowest-lying three-particle $N \pi \pi$
          energy levels. The thick solid black line marks the mass of the Roper resonance. 
	} 
	\label{fig:FitNsigma}
\end{figure}

Next, we include the dimer fields starting with the $\sigma$-dimer, which we studied in detail
throughout this work. We set $f_2$ and $f_3$ to zero, leaving us with the self-energy $\Sigma_{N \sigma}$
only. The numerical results for the $N \sigma$ contribution are displayed in Fig.~\ref{fig:FitNsigma}.  
In this system, the free, non-interacting three-particle $N \pi \pi$ energies are determined as 
\begin{align}
\begin{split}
E_{\pi \pi N}^{\text{free}} \left( \vec{n}_1 , \vec{n}_2 , \vec{n}_3 \right) &= \sqrt{m_N^2 + \left( \frac{2 \pi}{L} \right)^2 | \vec{n}_1 |^2 } 
+ \sqrt{M_{\pi}^2
+ \left( \frac{2 \pi}{L} \right)^2 | \vec{n}_2 |^2 } + \sqrt{M_{\pi}^2 + \left( \frac{2 \pi}{L} \right)^2 | \vec{n}_3 |^2 } \; . \label{pipiNfree-levels} 
\end{split}
\end{align}
There are, naturally, more free energy levels in this three-particle system, but some of them overlap with
each other. Also, it should be noted that not all possible combinations of the free $N \pi \pi$
system have the  quantum numbers of the Roper resonance $L_{2J2I}=P_{11}$. 
Since we did not include isospin, spin and angular momentum structures in our fundamental Lagrangian,
we simply show all interacting energy levels that appear in our calculation. In Fig.~\ref{fig:FitNsigma} 
the lowest lying free $N \pi \pi$ levels $N(0) \pi(0) \pi(0)$ (the $N \pi \pi$ threshold), $N(1) \pi(1) \pi(0)$,
$N(0) \pi(1) \pi(1)$, and $N(2) \pi(2) \pi(0)$ are shown. We observe that all our obtained energy levels
lie very close to the non-interacting three-particle levels and converge to them for large box sizes,
similar to the two-particle case from Fig.~\ref{fig:FitNpi}. The energy shift is negative caused by setting
$\alpha_{\sigma} = -1$ for the $\sigma$-dimer field. We tested what happens in the case that
$\alpha_{\sigma} = +1$ and, indeed, the interacting levels then approach the free levels from above.
There are no clear signs of avoided level crossing near the critical value. Solely the behaviour of the
energy level between the free levels $N(0) \pi(1) \pi(1)$ and $N(2) \pi(2) \pi(0)$ may be
affected by avoided level crossing, being first closer to $N(0) \pi(1) \pi(1)$, but then approaching
$N(2) \pi(2) \pi(0)$ for $M_{\pi} L > 5$. A possible explanation why no other signs of avoided level
crossing are visible might be the fact that the interacting energy levels lie too close to the free
levels, which can mitigate the typical signature of avoided level crossing. We tested that an increase
of the constants $h_2$ and $f_4$ within reasonable limits does not change this picture significantly.
In future studies, one should reconsider the numerical estimates of all involved LECs, perhaps with the
help of newly acquired lattice data.

\subsection{{\boldmath$\Delta\pi$} channel}
\begin{figure}[t]
	\begin{center}  
		\includegraphics[width=1.0\linewidth]{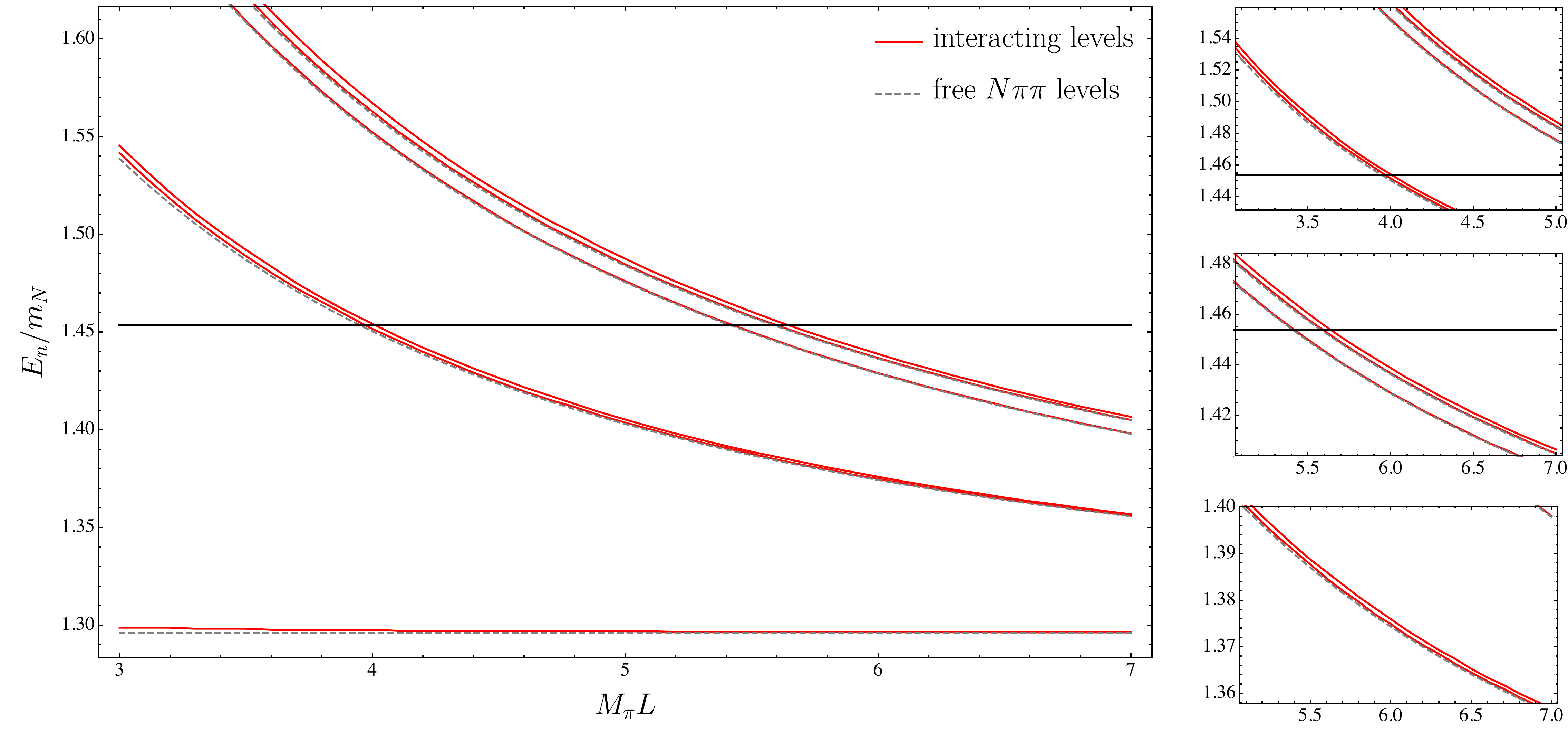}
	\end{center}
	\caption{
	  Energy levels for different box sizes $L$ considering only pion and $\Delta$-dimer as intermediate
          states. Red solid lines display the numerical results for the interacting energy levels and grey
          dashed lines the free (non-interacting) lowest-lying three-particle $N \pi \pi$ energy levels.
          The thick solid black line marks the mass of the Roper resonance. The small pictures on the
          right-hand side show more precisely the behaviour of the close lying energy levels.
	} 
	\label{fig:FitDeltapi}
\end{figure}

Now, we take a look at the second dimer-field, the $\Delta$-dimer. Analogously to the cases before, we set
the LECs $f_2$ and $f_4$ to zero, leaving us with the self-energy contribution $\Sigma_{\Delta \pi}$ only.
The results are shown in Fig.~\ref{fig:FitDeltapi}. 
Like in the $\sigma$-dimer spectrum, the obtained energy levels lie very close to the non-interacting
levels and asymptotically approach them for larger box sizes. This time the free levels are approached
from above due to $\alpha_{\Delta} = +1$ and the distance between the interacting and non-interacting levels is
overall much smaller than in the $N \sigma$ case. Also, in Fig.~\ref{fig:FitDeltapi} there are no visible signs of avoided level crossing.
Instead, another interesting effect appears in this spectrum: Above the free levels $N(1) \pi(1) \pi(0)$
and $N(2) \pi(2) \pi(0)$ there are two interacting energy levels visible, which lie very close, but
do not cross each other when increasing $M_\pi L$, see the zoom-in in Fig.~\ref{fig:FitDeltapi}. Indeed,
these energy levels belong to the same free energy eigenvalue, i.e. the lower energy double line
belongs to $N(1) \pi(1) \pi(0)$ and the upper one to $N(2) \pi(2) \pi(0)$. 
We tested this by reducing the coupling $g_2$, which causes both double lines to move closer to their
respective free energy levels and also decreases the splitting between the levels. The splitting
of these interacting energy levels comes from the fact that in the $\Delta \pi$ system either a
spectator pion or a pion within the $\Delta$-dimer propagator ($\Delta\to\pi N\to\Delta$) can
carry momentum away. Since both possibilities come with a different LEC, $f_3$ or $g_2$, respectively,
there is a small splitting between the levels.
This also explains why we did not see such a splitting of the interacting levels in the $N \sigma$ spectrum.
There, the nucleon is the spectator particle and the two pions interact with each other in the $\sigma$-dimer
propagator, so that it does not matter which pion carries away the momentum.
The question whether this splitting should be observed in a full coupled-channel ($\pi N/\Delta\pi/\sigma N$)
calculation brings us to an interesting point. In particular, a coupled $\Delta\pi/\sigma N$ system
allows for the appearance of a (pion) exchange diagram. These exchange diagrams enable transitions
between $\Delta$- and $\sigma$-dimer fields, which are important to fulfill unitarity. Such contributions,
however, cannot be included at leading one-loop order in the self-energy, but enter at two-loop order.
This issue is left out for a future work.

\subsection{{\boldmath$N\pi/N\sigma$} coupled-channel}
\begin{figure}[t]
	\begin{center}  
		\includegraphics[width=1.0\linewidth]{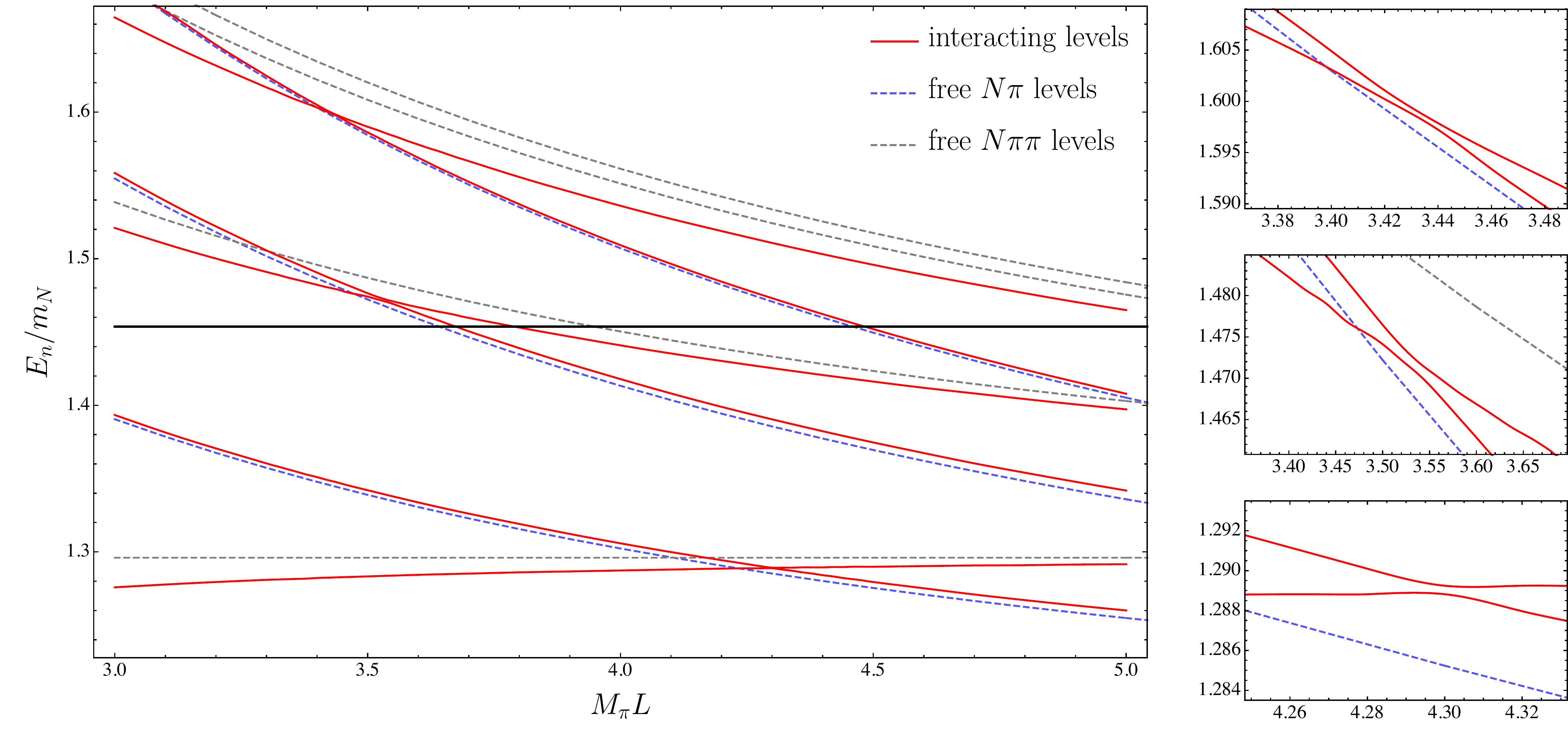}
	\end{center}
	\caption{
	  Energy levels for different box sizes $L$ considering the coupled channel with $N \pi$ and
          $N \sigma$ self-energy contributions. Red lines display the numerical results for the interacting
          energy levels. Blue dashed and grey dashed lines show the non-interacting $N \pi$ and $N \pi \pi$
          energy levels, respectively. The thick solid black line marks the mass of the Roper resonance.
          The small pictures on the right-hand side show the three critical points where the interacting
          energy levels come very close to each other.
	} 
	\label{fig:CoupledChannelSigma}
\end{figure}

For our final analysis we take a look at a coupled $N\pi/N\sigma$ system. This means that we include
both self-energy contributions at once in Eq.~\eqref{eq:Energy-levels-FV-final}, neglecting only
the $\pi\Delta$ ($f_3 = 0$) part for the reasons discussed before. 
The results of the coupled-channel energy levels are depicted in Fig.~\ref{fig:CoupledChannelSigma}. 
We restricted ourselves to $M_{\pi} L\le 5$, since many energy levels appear in this case, many
of which lie too close to the non-interacting ones.
Furthermore, we note that the free levels can naturally cross as a function of $M_\pi L$, see the grey
lines in Fig.~\ref{fig:CoupledChannelSigma}. However, a crossing of interacting levels would be in conflict
with the hermiticity of the perturbation theory Hamiltonian~\cite{Jordan:1933vh}. Indeed, this does not occur
as shown in the close-ups on the right-hand side of Fig.~\ref{fig:CoupledChannelSigma}. Furthermore, we
observe that the avoided level crossing signature of the two-particle $N \pi$ spectrum seen
in Fig.~\ref{fig:FitNpi} is now washed out in the coupled channel case, i.e. the interacting levels now
lie much closer to the free energy levels for small $M_{\pi} L$. This is probably caused by the
large contribution from the double sum in the $N \sigma$ self-energy contribution, which gives the
whole self-energy function an offset, that pushes the zeros of the function (interacting energy levels)
closer to its poles (non-interacting energy levels).

\subsection{Comparison to lattice QCD results}
\begin{figure}[t]
	\begin{center}  
		\includegraphics[width=0.8\linewidth]{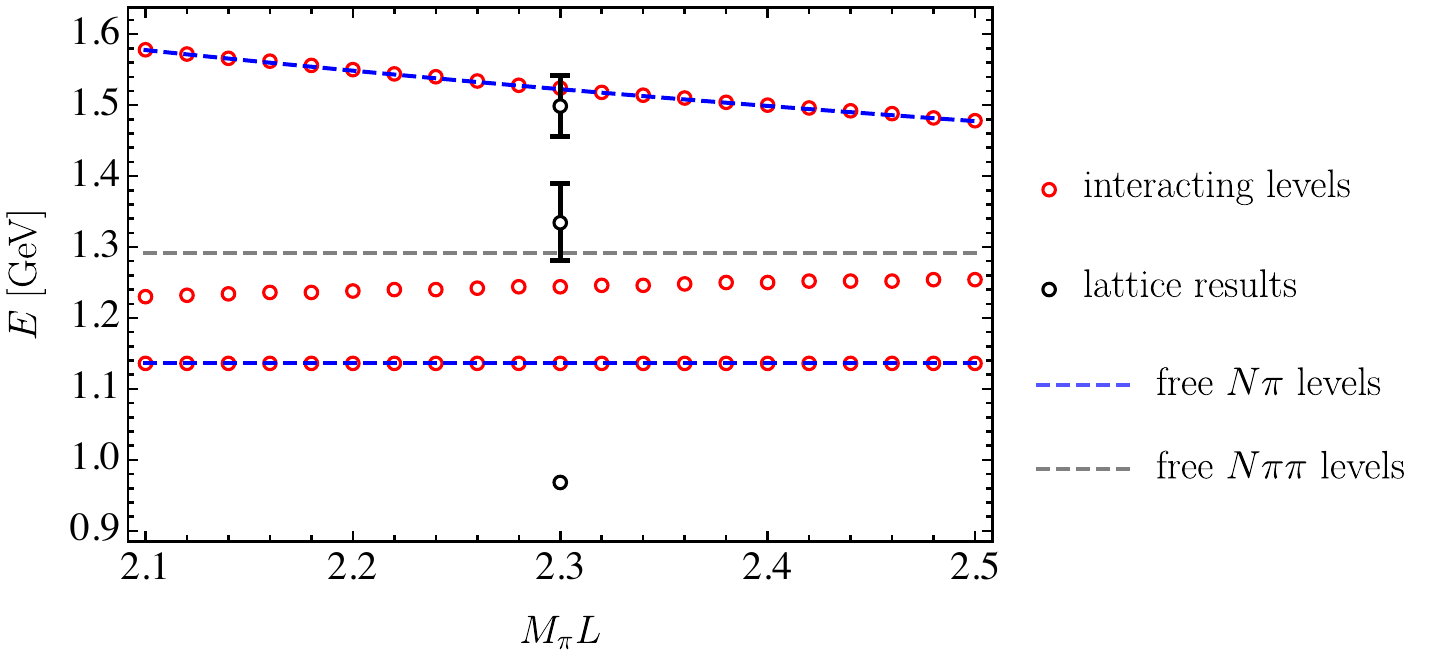}
	\end{center}
	\caption{
	  Comparison of the Roper resonance energy levels with lattice results using the $N \pi$ and $N \sigma$
          self-energy contributions. Red circles display the numerical results for the interacting energy levels
          and black circles the lattice results with errors from Ref.~\cite{Lang:2016hnn}. Blue dashed and grey
          dashed lines show the non-interacting $N \pi$ and $N \pi \pi$ energy levels, respectively. 
	} 
	\label{fig:LatticecompareSigma}
\end{figure}

Lastly, we can test how our results compare to previously obtained lattice QCD results from Ref.~\cite{Lang:2016hnn}.
Therein, the energy eigenvalues in the $G_1^+$ irreducible representation have been obtained in a
box of length $M_{\pi} L = 2.3$ with a pion mass close to the physical point, i.e. $M_{\pi} = 156\,$MeV, and a nucleon mass of $m_N \approx 980\,$MeV, also slightly larger than the physical value. To ensure a better
comparison with the lattice results, we use these values for $M_{\pi}$ and $m_N$.
The other masses and LECs in our calculation are not changed, i.e. we use the same estimates as
described before in section~\ref{sec:Numerics}. The comparison of our $N\pi/N\sigma$ coupled channel and the lattice results is shown in Fig.~\ref{fig:LatticecompareSigma}. 
We observe that the lattice QCD study found an energy level located at the nucleon mass, since the nucleon
has the same quantum numbers as the Roper resonance. In our calculation, this nucleon energy level does
not exist, because there is no self-energy contribution that produces a nucleon pole. Instead, our ground-state
level is located at the $N \pi$ threshold which, however, does not have the correct quantum numbers. The $N \pi$ threshold has negative parity meaning that it cannot show up in the Roper channel. Still, since no projection to definite parity is done here, this state appears as the lowest level in the $N \pi$ self-energy contribution from Eq.~\eqref{Sigma-N-pi-FV-correction}. Note that in the baryon chiral perturbation theory framework of Ref.~\cite{Severt:2020jzc} the $N \pi$ threshold does not appear since the chiral effective Lagrangian with all the proper symmetries forbids this state. Hence, the appearance of this threshold can be seen as an artifact of our non-relativistic EFT approximation. Once our formalism here is extended to include more symmetries and structures from chiral effective Lagrangians, we expect that the $N\pi$ threshold does not enter the spectrum anymore. 
The next higher energy level is the $N \pi \pi $ threshold. Our prediction for the corresponding interacting
energy level lies slightly below the threshold, whereas the lattice prediction lies just above it.
The error of the lattice result, however, is large enough to also allow a level below the threshold.
The next observed level corresponds to the first momentum including free level, i.e. $N(1) \pi (1)$.
Here, our prediction lies barely above the free level, but agrees with the lattice results within
the $1\sigma$ uncertainty quoted there~\cite{Lang:2016hnn}.

\begin{figure}[t]
	\begin{center}  
		\includegraphics[width=0.8\linewidth]{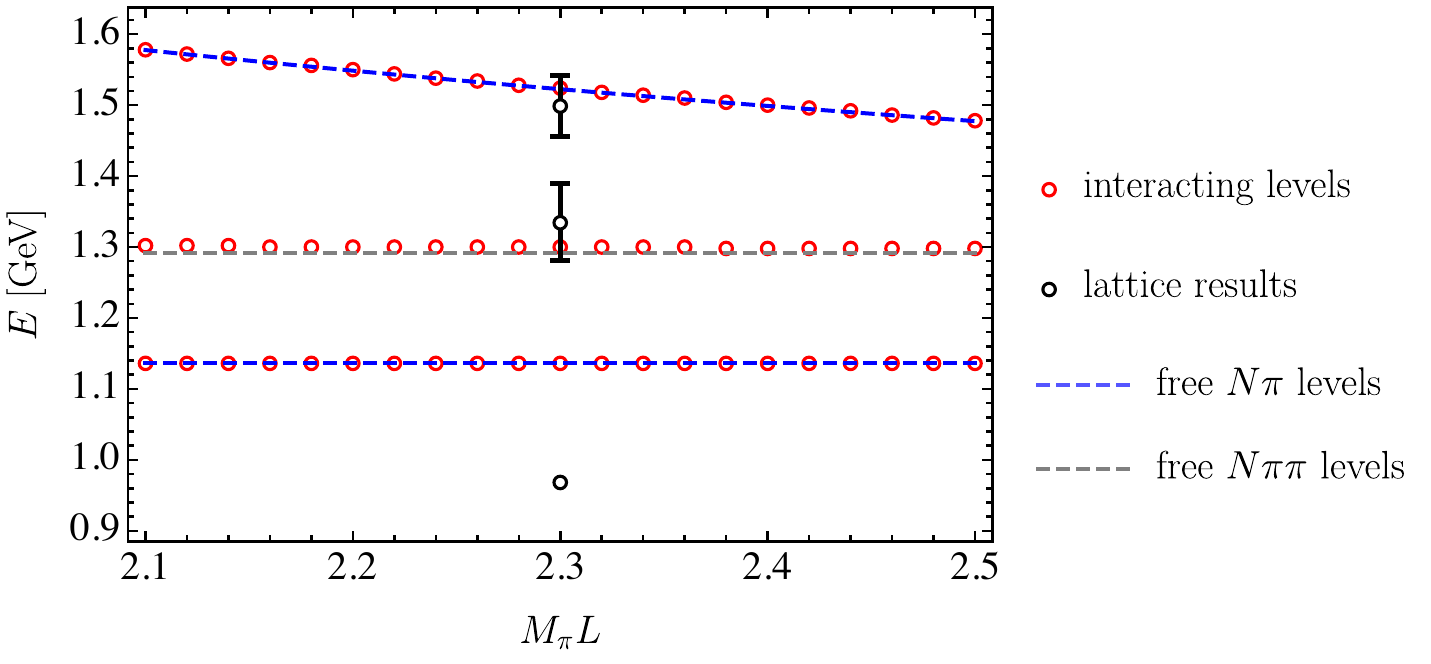}
	\end{center}
	\caption{
	  Comparison of the Roper resonance energy levels with lattice results using the $N \pi$ and $\Delta \pi$
          self-energy contributions. Red circles display the numerical results for the interacting energy levels
          and black circles the lattice results with errors from Ref.~\cite{Lang:2016hnn}. Blue dashed and grey
          dashed lines show the non-interacting $N \pi$ and $N \pi \pi$ energy levels, respectively.
	} 
	\label{fig:LatticecompareDelta}
\end{figure}

For completeness, we also consider the $N \pi/\Delta \pi$ coupled-channel for the comparison with the
lattice results. Setting $f_4 = 0$ and turning on the $\Delta$-dimer contribution, the finite-volume spectrum
is obtained and depicted in Fig.~\ref{fig:LatticecompareDelta}. 
The spectra look almost identical to Fig.~\ref{fig:LatticecompareSigma}. We again include the $N \pi$ threshold in the figure according to our explanation from before. The only difference is that the prediction
related to the $N \pi \pi$ threshold lies now slightly above the free level, which actually creates a better
overlap with the lattice result but also makes our prediction more consistent with a non-interacting theory.  
However, more work is needed here to find a suitable way to include both
$\Delta$-dimer and $\sigma$-dimer fields in one coupled channel. Also, for both plots,
Figs.~\ref{fig:LatticecompareSigma} and~\ref{fig:LatticecompareDelta}, we emphasize that the box length
is relatively small with $M_{\pi} L = 2.3$ meaning that exponentially suppressed contributions can still
give sizeable corrections at this point. Some of these contributions have been neglected in our finite-volume
approach, which can lead to further systematic uncertainties.

Nevertheless, we see that even without fitting to the lattice energy eigenvalues and assuming that the other parameters (masses and LECs) do not change by increasing the pion and nucleon mass, our predictions agree well with the lowest-lying states of the lattice spectrum. More specifically, we observe that our highest energy-eigenvalue ($\sim 1.6~{\rm GeV}$) is barely shifted from the corresponding free-energy irrespectively to the inclusion of $N\sigma$ or $\Delta\pi$ fields. The next lower energy-eigenvalue is shifted down/up from the $N\pi\pi$ free-energy, respectively to the $\{N\pi,N\sigma\}$ or $\{N\pi,\Delta\pi\}$ cases. Neither of these cases can be preferred statistically from the currently available lattice QCD results. Still, the fact that the energy shift from the free energy has different signs when including $N\sigma$ or $\Delta\pi$ cases tells one that when higher precision lattice results are available we indeed have the chance to resolve interaction patterns of the Roper. 


\section{Summary and conclusions}
\label{sec:summary} 

In this paper, we have analyzed the finite-volume spectrum of the Roper resonance using a particle-dimer
approach. We introduced a non-relativistic covariant Lagrangian with nucleons, pions and three dimer fields as degrees of freedom. These dimer fields are the Roper resonance itself, the
$\sigma$-meson and the $\Delta$-resonance. We then calculated the Roper self-energy within our framework
to one-loop order. Furthermore, we analyzed the $\sigma$- and $\Delta$-dimer fields and dressed
their corresponding propagators to explicitly include three-particle dynamics. From then on, we
restricted ourselves to a finite volume. We showed how the self-energy of the Roper resonance can
be calculated in a finite volume and how to extract the interacting energy levels of the Roper
system. Afterwards, we discussed methods to determine the appearing LECs that contribute to
the self-energy corrections. Then, we calculated the finite-volume spectra of the Roper resonance
for various cases. Our main findings are the following: 
\begin{itemize}
\item In the $N \pi$ channel, avoided level crossing can clearly be observed around the Roper resonance mass.
  For large box sizes, the energy levels approach the free $N \pi$ energies. The spectrum agrees very well
  with our previous result in Ref.~\cite{Severt:2020jzc}, using baryon chiral perturbation theory. 
\item Including the $N\sigma$ channel, with the $\sigma$ dressed by the pertinent $\pi\pi$ loops, we were
  able to implement 
  three-body ($N \pi \pi$) dynamics. While we checked that the two-body sub-system can reproduce the 
  finite-volume spectrum for not too large pion masses, no clear signs of avoided level crossing could
  be observed in the three-body ($N\pi\pi$) spectrum. We observed similar behaviour for the $\Delta\pi$ channel.
\item Uniting the $N \pi$ and $N \sigma$ contributions in a coupled-channel system, we observed that
  the interacting energy levels lie very close to their respective free $N \pi$ or $N \pi \pi$ levels. Strikingly, the obtained spectrum in our formalism showed an overall good agreement to the lattice QCD results~\cite{Lang:2016hnn} even without a fit to their energy eigenvalues. 
\end{itemize}
In conclusion, we think that albeit very simple, the proposed alternative finite-volume formalism defines
a new, systematically improvable pathway of extracting resonance proper-ties from finite-volume spectra. Moreover, already now the formalism shows that effects due to $N\sigma$ and $\Delta\pi$ channels can be decomposed once more precise lattice results are available. With that, the formalism provides already at this stage a valuable guidance on the required precision of the lattice QCD input.
Systematical updates to the formalism include spin and isospin projections as well
as inter-couplings between different particle-dimer channels via pion exchange diagrams, so that a full $N \pi / N \sigma / \Delta \pi$ coupled-channel system can be achieved. Work in this direction is planed. 

\acknowledgments
We thank A. Rusetsky and F. M{\"u}ller for many useful discussions. This work is supported by the
Deutsche Forschungsgemeinschaft (DFG, German Research Foundation), the NSFC through the funds provided to
the Sino-German Collaborative Research Center CRC 110 “Symmetries and the Emergence of Structure in QCD”
(DFG Project-ID 196253076 - TRR 110, NSFC Grant No.~12070131001). The work of UGM was
further supported by VolkswagenStiftung (grant No. 93562) and by the Chinese Academy of Sciences
(PIFI grant 2018DM0034). MM was further supported by the National Science Foundation under Grant No. PHY-2012289.


\bibliographystyle{JHEP} 
\bibliography{BIB}

\end{document}